\documentstyle[aps,amstex,psfig,float]{revtex}

\def\n{\textbf{n}}
\def\r{\textbf{r}}
\def\x{\textbf{x}}
\def\y{\textbf{y}}
\def\z{\textbf{z}}
\def\k{\textbf{k}}

\def\K{\textbf{K}}
\def\f{\textbf{f}}
\def\t{\textbf{t}}
\def\G{{\cal G}}
\def\H{{\cal H}}
\def\A{{\cal A}}
\def\B{{\cal B}}
\def\d#1{\partial_{#1}}
\def\h{\frac{1}{2}}
\def\eq#1{Eq.~(\ref{#1})}
\def\fig#1{Fig.~\ref{#1}}

\def\tab#1{Tab.~\ref{#1}}
\def\sec#1{Sec.~\ref{#1}}
\def\app#1{Appendix~\ref{#1}}

\begin{document}

\title{A systematic approach to bicontinuous cubic phases in ternary 
amphiphilic systems}
\author{U. S. Schwarz\cite{weizmann} and G. Gompper}
\address{Max-Planck-Institut f\"ur Kolloid- und Grenzfl\"achenforschung, \\
Kantstr.~55, 14513 Teltow, Germany.} 
\date{\today}
\maketitle

\begin{abstract}
The Fourier approach and theories of space groups and color symmetries
are used to systematically generate and compare bicontinuous cubic
structures in the framework of a Ginzburg-Landau model for ternary
amphiphilic systems.  Both single and double structures are
investigated; they correspond to systems with one or two monolayers in
a unit cell, respectively.  We show how and why single structures can
be made to approach triply periodic minimal surfaces very closely, and
give improved nodal approximations for G, D, I-WP and P surfaces.  We
demonstrate that the relative stability of the single structures can be
calculated from the geometrical properties of their interfaces only.
The single gyroid G turns out to be the most stable bicontinuous cubic
phase since it has the smallest porosity.  The representations are
used to calculate distributions of the Gaussian curvature and
$^2H$-NMR bandshapes for C(P), C(D), S, C(Y) and F-RD surfaces.
\end{abstract}


\section{Introduction}
\label{sec:intro}

\par Amphiphiles are molecules which have both hydrophilic and
hydrophobic parts.  In a ternary mixture of amphiphiles, water and
oil, their ability to stabilize water-oil interfaces leads to
structure formation on the nanometer scale
\cite{a:gelb94,a:Gomp94,a:lipo95c}.  Depending on concentration and
temperature, many different phases are found to be stable.  One
intriguing aspect of amphiphilic polymorphism is the existence of
ordered bicontinuous phases which can be traversed in any direction in
both the water and the oil regions.  Here macroscopic order has been
demonstrated by the appearance of Bragg-peaks in diffraction patterns
\cite{a:auss90} and bicontinuity by measuring diffusion properties
with nuclear magnetic resonance \cite{a:ande90b}.  Clearly any
bicontinuous ordered phase has to have a three-dimensional
Bravais-lattice.  In experiments, cubic symmetry is observed in most
cases \cite{a:font90,a:luzz93,a:sedd95}.  The amphiphilic monolayers of
bicontinuous cubic structures, which separate regions with water from
regions with oil, are often modelled by triply periodic surfaces of
constant mean curvature \cite{a:ande88,a:wang90}.  By changing
experimental parameters like temperature or salt concentration, their
preference to bend towards the water or the oil regions can be made to
vanish.  Then one can describe them by triply periodic minimal
surfaces (TPMS).  Similarly, the bilayers of bicontinuous cubic phases
in lipid-water mixtures by symmetry have no spontaneous curvature, so
that their mid-surfaces can also be modelled by TPMS
\cite{a:hyde97}.  Bicontinuous cubic phases have gained
renewed interest recently, for example as space partitioners in
biological systems \cite{a:land95c}, as amphiphilic templates for
mesoporous systems \cite{n:atta95} and for the crystallization of
membrane proteins \cite{a:peba97}.

\par In this paper, we report on a systematic theoretical
investigation of bicontinuous cubic phases in ternary amphiphilic
systems with vanishing spontaneous curvature.  We restrict ourselves
to bicontinuous ordered phases with cubic symmetry since these are the
ones which are mainly observed in experiments. It has been shown
recently by G{\'o}{\'z}d{\'z} and Ho{\l}yst \cite{a:gozd96b} in the
framework of real-space minimization that various cubic bicontinuous
phases can be generated as local minima of the Ginzburg-Landau model
for ternary amphiphilic systems introduced previously by Gompper and
Schick \cite{a:gomp90,a:Gomp94}. The position of the interfaces
between oil and water regions then follow as iso-surfaces to a scalar
field $\Phi(\r)$. Our analysis is based on the Fourier ansatz in
combination with the theories of space groups and color symmetries; it
allows to generate the relevant structures in a very systematic way
and to reduce the representations to a relatively small number of
variables.  This in turn allows very efficient numerics and makes it
easy to document and reuse the results.

\par In order to analyze our results, we use an effective interface
Hamiltonian, which is derived from the Ginzburg-Landau theory
\cite{a:gomp92a}.  Due to the oil-water symmetry of the
Ginzburg-Landau model, the spontaneous curvature vanishes.  Since the
elastic energy of interfaces is governed by their bending rigidity,
the interfaces of the structures which have only one amphiphilic
monolayer are very close to cubic TPMS.  A similar approach, which
employs a Ginzburg-Landau model for binary fluid mixtures, has been
used in Refs.~\cite{a:shen94a,a:shen94}.  However, since the elastic
energy of interfaces is governed by their tension in this case, the
numerical minimization is much less stable.  In fact, our tabulation
of the first few Fourier amplitudes for the four best-known structures
G, D, I-WP and P provides a considerable improvement over the widely
used nodal approximations for TPMS \cite{a:schn91}.  Since most of the
detailed analysis of the properties of cubic TPMS has been focused on
these structures so far, we want to use our representations to
characterize the structures C(P), C(D), S, C(Y) and F-RD in more
detail.

\par Our analysis in terms of interfacial properties also allows us to
investigate the relative stability of the various bicontinuous cubic
phases in terms of their geometrical properties.  Although the
identification of a certain bicontinuous cubic structure is a tedious
procedure best accomplished by combining several experimental
techniques like neutron or X-ray small angle scattering, transmission
electron microscopy and nuclear magnetic resonance, in most cases one
has found the stable phases to be either gyroid or diamond
structures. Here we derive an universal geometrical criterion for
their relative stability in the case of vanishing spontaneous
curvature of the interfaces which explains these experimental findings
and is in excellent agreement with our numerical results.

\par The paper is organized as follows.  We begin by introducing the
relevant concepts from differential geometry and crystallography and
their applications to amphiphilic systems in \sec{sec:geometry}.  In
\sec{sec:fourier} we explain how to construct Fourier series and nodal
approximations for amphiphilic structures.  In \sec{sec:phi6} we
systematically generate, evaluate and compare various single and
double structures by using the Ginzburg-Landau theory, as well as the
interface Hamiltonian which follows from it.  In
\sec{sec:experimental} we use the representations obtained to comment
on the experimental identification of bicontinuous cubic phases in
amphiphilic and mesoporous systems.

\section{Geometry of bicontinuous phases}
\label{sec:geometry}

\par If we characterize a bicontinuous phase by its interfaces, the
simplest possible case is that they form \emph{one} triply periodic
surface (TPS), i.e.\ a surface with a three-dimensional
Bravais-lattice.  Any TPS divides space into two unconnected but
intertwined labyrinths.  In ternary amphiphilic systems, the surface
is covered with an amphiphilic monolayer and the two labyrinths are
filled with water and oil, respectively.  Therefore any structure
corresponding to a TPS is bicontinuous in water and oil.  The TPS is
called \emph{balanced} if there exists an Euclidean transformation
$\alpha$ which maps one labyrinth onto the other.  Oil and water have
equal volume fractions in this case.

\par In the mathematical context, triply periodic \emph{minimal}
surfaces (TPMS) have been investigated extensively
\cite{a:nits89,a:dier92}.  In order to locally minimize their surface
area, minimal surfaces have vanishing mean curvature $H = ( 1/R_1 +
1/R_2 ) / 2$ at every point on the surface, where $R_1$ and $R_2$ are
the two principal radii of curvature.  Therefore, their Gaussian
curvature $K = 1/R_1 R_2 = - 1/R_1^2$ is non-positive and the surface
is saddle-shaped everywhere.  Up to 1970 only five TPMS had been known
by the 19th-century work of Schwarz and his students, with P, D and
C(P) being cubic.  Then Schoen described 12 more \cite{a:scho70},
including the cubic cases G, F-RD, I-WP, O,C-TO and C(D).  Their
existence was proven by Karcher in 1989
\cite{a:karc89}.  More TPMS have been found by Karcher and Polthier
\cite{a:karc96}, Fischer and Koch \cite{a:fisc96} and others, but all
of them seem to be more complicated than the ones known to Schoen.
For cubic TPMS, exact representations are known only for P, D, G and
I-WP.  These \emph{Weierstrass representations} originate from
complex analysis and are explained in \app{sec:weierstrass}.  With
their help the surface areas of these TPMS can be calculated exactly
in terms of elliptic functions; the resulting values are given in
\tab{WeierstrassTabelle}.

\par Being balanced and being minimal are two independent possible
properties of a TPS.  From the cubic TPMS known to Schoen, F-RD, I-WP
and O,C-TO are not balanced and divide space into two labyrinths with
unequal volume fractions.  On the other hand, the TPS defined by $\cos
x + \cos y + \cos z = 0$ is balanced with $\alpha$ being a translation
along half the body diagonal of the conventional unit cell, but far
from minimal.  If a TPS contains straight lines, it is called
\emph{spanning} \cite{a:fisc87}.  It was shown by Schwarz that any
spanning TPMS is symmetric with respect to a reflection around such a
line.  Since this operation exchanges the two labyrinths, any spanning
TPMS is balanced.  The only known cubic TPMS, which is balanced but
not spanning, is the gyroid G, where the symmetry operation exchanging
the two labyrinths is the inversion.

\par For our purposes it is best to investigate not the local
concentrations of oil and water separately, but their difference
$\Phi(\r) = \rho_O(\r) - \rho_W(\r)$.  Since we assume
incompressibility, $\rho_O(\r) + \rho_W(\r) = 1$ and the local
concentrations of oil and water can be reconstructed from $\Phi(\r)$.
The amphiphilic monolayers now correspond to the $\Phi=0$
iso-surfaces.  The interchange of oil and water amounts to $\Phi \to -
\Phi$.  For balanced surfaces this is a symmetry operation when
combined with the Euclidean operation $\alpha$, that is $\Phi(\alpha
\r) = - \Phi(\r)$.  However, it is important to note that this is not
a space group symmetry --- which would only allow for a rigid motion,
but not for a change of sign.  Additional symmetries which arise when
an Euclidean operation is combined with a permutation are called
\emph{color symmetries}; with only water and oil present we have the
simplest case of a \emph{black and white symmetry}
\cite{a:schw84,a:sene90,a:lifs97}.  When water and oil are colored
white and black, respectively, the $\Phi=0$ iso-surface obtains
differently colored sides and thus can be considered to be oriented.
We assume the uncolored structure (i.e.\ the unoriented surface) to
have space group $\G$.  Then all Euclidean operations which do not
interchange the two labyrinths form a subgroup $\H$ of $\G$ of index
$2$.  By definition $\H$ is the space group of the colored structure
(i.e.\ the oriented surface).  The quotient group is isomorphic to the
cyclic group of order $2$, that is $\G / \H \cong {\Bbb Z}_2 \cong \{ 1,
\alpha \}$.  In general, any balanced TPS is characterized by a
group-subgroup pair $\G - \H$ of index $2$.  For example, for the
balanced TPS defined by $\cos x + \cos y + \cos z = 0$, we have $\H =
Pm\bar3m$ (No.~221) and $\G = Im\bar3m$ (No.~229).

\par
\begin{figure}[H]
\begin{center}
\begin{tabular}{cc}
\psfig{file=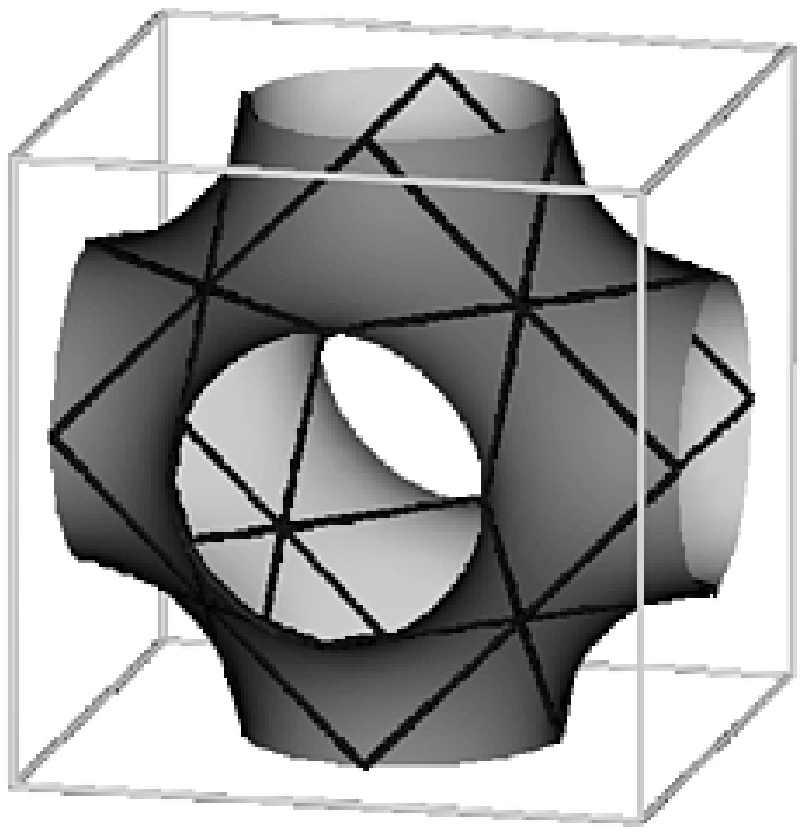,width=7cm} &
\psfig{file=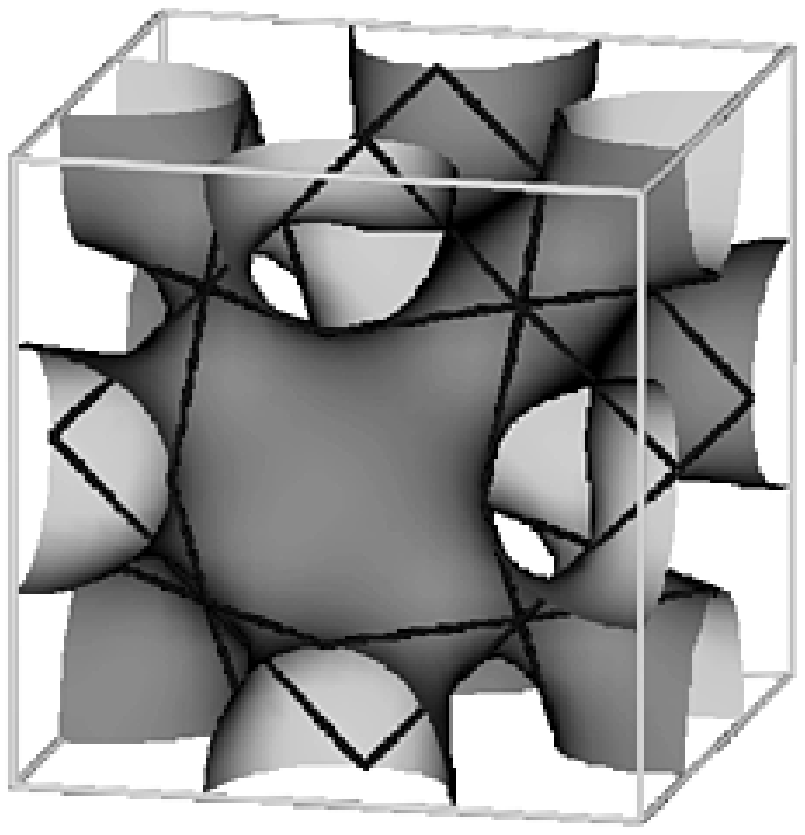,width=7cm} \\
P & C(P) \\
\psfig{file=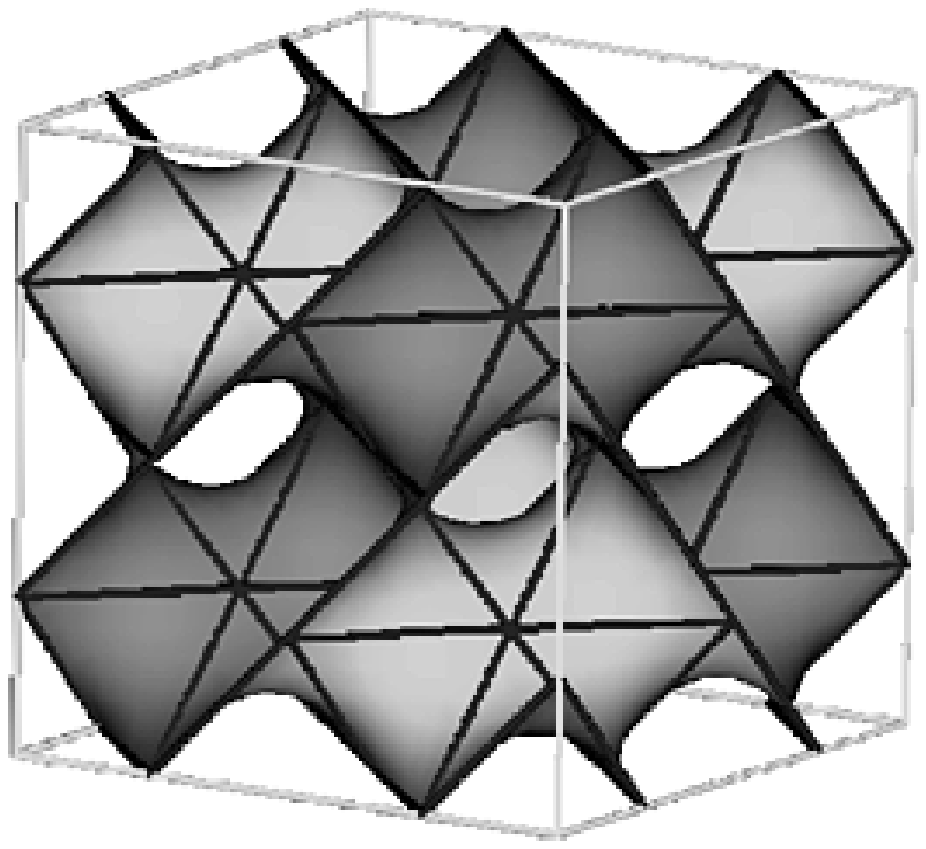,width=7cm} &
\psfig{file=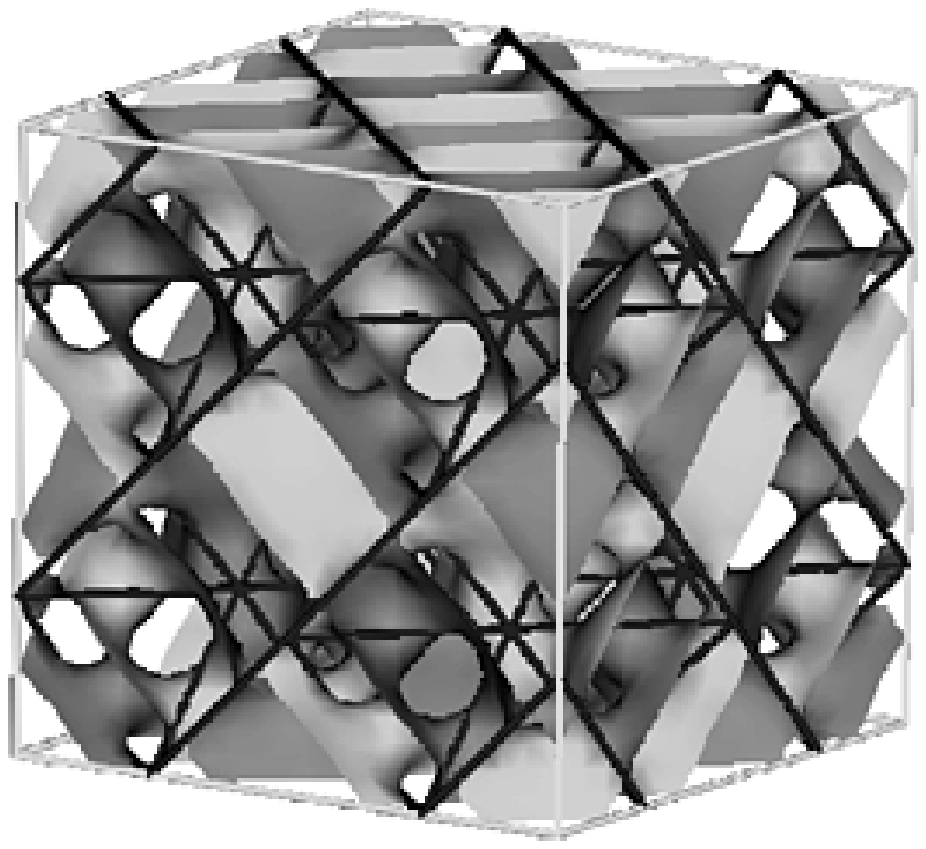,width=7cm} \\
D & C(D) \\
\end{tabular}
\end{center}
\caption{Different TPMS with the same space group and the same system
of straight lines.  Spanning minimal surfaces are always balanced, 
thus they are characterized by a group-subgroup pair $\G - \H$.  For P and
C(P) this is $Im\bar{3}m$ (No.~229) - $Pm\bar3m$ (No.~221) and for D
and C(D) this is $Pn\bar{3}m$ (No.~224) - $Fd\bar{3}m$ (No.~227).}
\label{StraightLines}
\end{figure}

 In order to describe a balanced TPS which does not intersect
itself, certain restrictions apply for the extension from $\H$
to $\G$ (for example, if $\alpha$ is a rotation, it has to be 2-fold).
This has been used by Fischer and Koch to develop a crystallographic
classification of all cubic TPMS \cite{a:fisc87,a:fisc96}.  In
particular they showed that only 34 cubic group-subgroup pairs $\G -
\H$ with index 2 can correspond to cubic balanced TPMS.  Although
there is no way to systematically list all balanced TPMS belonging to
a possible group-subgroup pairing $\G - \H$, Fischer and Koch argued
that one can completely enumerate all cubic spanning TPMS which have
straight lines forming a three-dimensional network whose polygons are
spanned disc-like by the surface.  They find P, C(P), D, C(D), S and
C(Y), with the latter two described by them for the first time.  In
our work we systematically generate these structures together with the
only known balanced but non-spanning cubic TPMS G and the non-balanced
ones I-WP and F-RD.  Examples for spanning TPMS are shown in
\fig{StraightLines}.

\par One also can consider structures with \emph{two} TPS.  For
example, parallel surfaces can be constructed on either side of any
given TPS.  Both parallel surfaces are TPS, each of which has two
labyrinths, which are topologically equivalent to those of the initial
structure.  However, with respect to the overall structure, these
labyrinths are now filled with the same component and separated by a
bilayer which is filled with the other component.  One might call this
structure \emph{tricontinuous}, but we rather prefer to use the term
\emph{double structures} for structures of \emph{two} TPS --- in
contrast to \emph{single structures} containing \emph{one} TPS.

\par Experimentally, little is known about the exact structure of
bicontinuous cubic phases in ternary amphiphilic systems.  One
exception is the system DDAB-water-styrene, for which five different
double structures have been found to be stable, each of which can be
regarded as an oil-filled bilayer draped onto a different TPS
\cite{a:stro92}.  However, with microemulsion phases being so
ubiquitous for these systems, single structures can be expected also
to be stable, and in fact have been reported in Ref.~\cite{a:raed89}.  
In contrast to the situation for (almost balanced)
ternary amphiphilic systems, the structure of bicontinuous cubic
phases is well-established for many binary amphiphilic systems
\cite{a:font90,a:luzz93,a:sedd95}.  Here, mostly inverse phases are
found, in which the two labyrinths are filled with water and the TPS
is covered with an amphiphilic bilayer.  These structures can be
considered to be double structures of a ternary system where the oil
has been removed from the bilayer.  However, if the thickness of the
bilayer is small compared to the lattice constant, it seems more
appropriate to model them by single structures of a ternary system,
where the water regions on both sides of the bilayer are distinguished
artificially by labelling them `water I' and `water II', respectively
\cite{a:cate88}.  Thus our results for single structures can also be
applied to bicontinuous cubic phases in amphiphile-water mixtures.

\section{Fourier ansatz}
\label{sec:fourier}

\par In order to construct a Fourier ansatz for a given group-subgroup
pair $\G - \H$, we need a Fourier ansatz for space group $\H$, together
with the implementation of a certain black and white symmetry.  In
ternary amphiphilic systems, this corresponds to a balanced single
structure.  For non-balanced single structures and double structures
there is no color symmetry and one only needs the Fourier ansatz for a
given space group $\H \equiv \G$.

\par It is well known how to choose the Fourier ansatz for a given
space group $\H$ \cite{a:shmu96,a:merm92}.  In the cubic system the
translational properties of $\H$ correspond to one of the three cubic
Bravais lattices simple cubic (P), body centered cubic (I) or face
centered cubic (F).  The Fourier series then reads
\begin{equation} \label{Fourieransatz1}
\Phi(\r) = \sum_{\{ \K \}} \Phi_{\K}\ e^{i \K \cdot \r}\
\end{equation}
where $\{ \K \}$ is the reciprocal lattice and the $\Phi_{\K}$ are the
(complex) Fourier amplitudes. However, not all of them are
independent.  Since the density field $\Phi(\r)$ has to be real, it
follows from \eq{Fourieransatz1} that $\overline{\Phi_{\K}} =
\Phi_{-\K}$.  With $\Phi_{\K} = A_{\K} - i B_{\K}$ , 
\eq{Fourieransatz1} becomes
\begin{equation} \label{Fourieransatz2}
\Phi(\r) = \sum_{\{ \K \}} \left\{ A_{\K} \cos(\K \cdot \r) 
                                 + B_{\K} \sin(\K \cdot \r) \right\}\ .
\end{equation}
Here $A_{\K}$ and $B_{\K}$ are real and can be obtained from $\Phi(\r)$ 
as
\begin{equation} \label{FourierInvertiert}
A_{\K} = \frac{1}{v} \int d\r\ \Phi(\r)\ \cos(\K \cdot \r), \ \ \ \ \ \ 
B_{\K} = \frac{1}{v} \int d\r\ \Phi(\r)\ \sin(\K \cdot \r), 
\end{equation}
where the integration runs over some unit cell of volume $v$.  Thus,
for even or odd functions, $B_{\K} = 0$ or $A_{\K} = 0$, respectively.

\par Further restrictions on the Fourier amplitudes $\Phi_{\K}$ arise
when we consider symmetry operations $\r \to P \r + \t$, which are
not pure translations.  Here, $P$ denotes a rotation, inversion or
roto-inversion, and $\t$ a translation which does not belong to the
Bravais lattice.  Thus, $\t$ is non-vanishing for screw axes and glide
planes.  It now follows from \eq{Fourieransatz1} that $\Phi_{P^{-1}
\K} = \Phi_{\K}\ e^{i \K \cdot \t}$.  Let $H$ be the point group of
the given space group $\H$.  To each operation in $H$ there
corresponds a pair $(P, \t)$ in $\H$.  We group all reciprocal lattice
vectors, which are related by an operation of $H$, into so-called
\emph{Fourier stars}.  All members of a given star have the same
wavelength.  Only few stars will have the same wavelength and in most
cases there is only one star with a certain wavelength.  Therefore, the
stars can be ordered according to decreasing wavelength and numbered
with the index $i$.  For each star we arbitrarily choose one
representative $\K_i$.  With $\Phi_{\K_i} = A_{i} - i B_{i}$,
\eq{Fourieransatz1} becomes a sum over Fourier stars,
\begin{equation} \label{Fourieransatz3}
\Phi(\r) = \sum_i \frac{n_i}{n}\ \{ A_i\ \A_{\K_i}(\r) 
                                    + B_i\ \B_{\K_i}(\r) \} \ ,
\end{equation}
where the \emph{geometric structure factors} are defined by
\begin{eqnarray} \label{IT1}
\A_{\K_i}(\r) &=& \sum_{H} \cos[ (P^{-1} \K_i) \cdot \r + \K_i \cdot \t ], \\
\B_{\K_i}(\r) &=& \sum_{H} \sin[ (P^{-1} \K_i) \cdot \r + \K_i \cdot \t ]\ .
\end{eqnarray}
Their precise form is obtained by explicitly enumerating all
non-translational symmetry operations of space group $\H$. The results
of this procedure can be taken from Ref.~\cite{a:shmu96} if
they are corrected by a factor $1/m_L$, where $m_L$ is the
multiplicity of the Bravais lattice (1, 2 and 4 for P, I and F,
respectively).  In \eq{Fourieransatz3}, $n$ denotes the order of $H$
and $n_i$ the multiplicity of $\K_i$. If $H$ does not contain the
inversion, we can extend the definition of a Fourier star by using
$\overline{\Phi_{\K}} = \Phi_{-\K}$. This does not change
\eq{Fourieransatz3} except that now we have to replace $H$ by the
so-called \emph{Laue group}, that is $H$ extended by the inversion.
All structures investigated in this work have $m\bar3m$ with order $n
= 48$ either as their point group or as their Laue group, thus the
members of one star are generated by all possible permutations and
changes of sign of the components of $\K_i$.  With $A_i = A_{\K_i}$
and $B_i = B_{\K_i}$, every Fourier star has two variables.
This reduces to one when $\Phi$ is an even or odd function of
$\r$.

\par We now address the question how to implement the additional black
and white symmetry.  Since $\H$ has index $2$ in $\G$, it follows from
the theorem of Hermann \cite{a:sene90} that $\H$ has either the same
point group or the same Bravais lattice as $\G$.  If $\H$ and $\G$
have the {\it same point group}, their Bravais lattices have to be
different.  Thus the Euclidean operation $\alpha$, which maps oil
regions onto water regions, has to be a translation $\t_{\alpha}$
which transforms one cubic Bravais-lattice into another.  There
are only two possibilities:  for $\t_{\alpha} = a (\x + \y + \z) / 2$
a P-lattice becomes a I-lattice, and for $\t_{\alpha} = a \x / 2$ a
F-lattice becomes a P-lattice.  The operation $\Phi(\r + \t_{\alpha})
= - \Phi(\r)$ in combination with \eq{Fourieransatz1} leads to additional
reflection conditions:  $h+k+l = 2n+1$ for $P \to I$, and $h,k,l =
2n+1$ for $F \to P$.  Note that these reflection conditions are
complementary to the well-known ones for I and F ($h + k + l = 2n$ and
$h+k, h+l, k+l = 2n$, respectively).  Thus in the case of identical
point groups, the form of the Fourier stars in \eq{Fourieransatz3} for
space group $\H$ stays the same, but the sum now runs over a reduced
set of Fourier stars.

\par If $\H$ and $\G$ have the {\it same Bravais lattice}, their point
groups have to be different.  Thus the Euclidean operation $\alpha$
has to be a point group operation $P_{\alpha}$ which extends one cubic
point group into another one.  There are five cubic point groups and
six ways to extend one of them into another by some $P_{\alpha}$. From 
$\Phi(P_{\alpha} \r) = - \Phi(\r)$ and \eq{Fourieransatz1} we now
find $\Phi_{P_{\alpha}^{-1} \K} = - \Phi_{\K} \exp(i \K \cdot
\t_{\alpha})$.  Thus, different Fourier stars merge to form new ones.
However, in all cases considered in this work there is no need to
consider new Fourier stars, since we only deal with the simple case
that $P_{\alpha}$ is the inversion.  Then $\t_{\alpha} = \textbf{0}$
and one finds $A_i = 0$ for all $i$.  The same result of course
follows from \eq{Fourieransatz2}, since now $\Phi$ is an odd function
of $\r$.

\par \tab{nodal} lists the group-subgroup pairs $\G - \H$ and the
nature of the extension $\H \to \G$ for the nine single structures
investigated in this paper \cite{a:fisc87,a:fisc96}.  Out of
the seven balanced cases, five have identical point groups and two
have identical Bravais lattices with $P_{\alpha}$ being the inversion.
In the next section, we will discuss our numerical results for Fourier
series of functions $\Phi(\r)$, whose $\Phi = 0$ iso-surfaces
approximate triply periodic minimal surfaces.  However, a very simple
approximation can be obtained by only considering the space group
information.  The use of a single Fourier star leads to the so-called
\emph{nodal approximation}, which does not need any Fourier amplitude.
For balanced structures, it is essential to consider the correct
black-and-white symmetry.  If one Fourier star is not sufficient to
represent the topology correctly, one has to consider another one and
adjust its Fourier amplitude by visual inspection.  In the last column
of \tab{nodal}, nodal approximations are given for the nine single
structures investigated.  They have been derived in the context of
triply periodic zero potential surfaces of ionic crystals
\cite{a:schn91} and are widely used as approximations for TPMS (e.g.\
in Refs.~\cite{a:land95c,a:lamb96}).  However, it is well known
that their properties can differ considerably from those of real TPMS 
\cite{a:barn90}.  In the next section we will show that the
quality of the nodal approximation varies considerably for different
structures.  

\section{Triply periodic structures in Ginzburg-Landau models}
\label{sec:phi6}

\subsection{Ginzburg-Landau model}
\label{subsec:phi6model}

\par The simplest Ginzburg-Landau model \cite{a:gomp90}, which 
successfully describes the phase behavior and mesoscopic structure of
ternary amphiphilic systems, contains a single, scalar order parameter 
field $\Phi(\r)$; thus, amphiphilic degrees of freedom are integrated
out.  The model is defined by the free-energy functional
\begin{equation} \label{Phi6FunktionalOrt}
{\cal F}[\Phi] = \int\ d\r\ \left\{ (\Delta \Phi)^{2} 
+ g(\Phi) (\nabla \Phi)^{2} + f(\Phi) \right\}\ .
\end{equation}
For ternary amphiphilic systems at the phase inversion temperature,
the free-energy functional has to be symmetrical in water and oil,
{\it i.e.}  invariant under the transformation $\Phi \to - \Phi$.  A
convenient form for the free-energy density and the structural
parameter of the homogeneous phases are \cite{a:gomp93a}
\begin{equation} \label{GL_density}
f(\Phi) = (\Phi + 1)^{2} (\Phi - 1)^{2} (\Phi^{2} + f_0)\ , \ \ \ \ \ \ 
g(\Phi) = g_0 + g_2 \Phi^{2}\ .
\end{equation}
Here, the three minima at $-1$, $0$ and $1$ correspond to excess
water, microemulsion and excess oil phases, respectively, and $f_0$
acts as a chemical potential for amphiphiles.  The behavior of the
correlation function $\langle \Phi({\bf r}_1) \Phi({\bf r}_2) \rangle$
is controlled by the parameters $g_0$ and $g_2$ in the microemulsion
and excess phases, respectively; they are therefore related to the
amphiphilic strength and the solubility of the amphiphile
\cite{a:Gomp94}.  Strongly structured microemulsions are obtained for
$g_0<0$, while unstructured excess phases require $(g_2+g_0)$ to be
positive and sufficiently large.

\par The elastic properties of amphiphilic monolayers are described 
by the \emph{Canham-Helfrich Hamiltonian} \cite{a:canh70,a:helf73}
\begin{equation} \label{HelfrichHamiltonian} 
{\cal H} = \int dA\ \left\{ \sigma + 2 \kappa\ (H - c_0)^2 
                                   + \bar \kappa\ K \right\}\ ,
\end{equation}
where the integration extends over the surface, $H$ and $K$ are mean
and Gaussian curvature, respectively, and the elastic moduli are the
surface tension $\sigma$, the spontaneous curvature $c_0$, the bending
rigidity $\kappa$ and the saddle-splay modulus $\bar \kappa$.  If the
$\Phi = 0$ iso-surfaces are identified with the location of the
amphiphilic monolayers, the free energy of the Ginzburg-Landau model
(\ref{Phi6FunktionalOrt}) can be well approximated by a sum of the
curvature energy --- calculated from \eq{HelfrichHamiltonian} --- and
a direct interaction between the interfaces, which falls off
exponentially with distance \cite{a:gomp92a,a:gomp93a}.  Due to the
water-oil symmetry the spontaneous curvature $c_0$ vanishes.  The
values of the other elastic moduli can be calculated numerically from
the profile $\Phi_s(z)$, which minimizes the free energy of a planar
water-oil interface \cite{a:gomp92a}.

\par With $f_0$, $g_0$ and $g_2$, the model has a three-dimensional
parameter space.  We investigate bicontinuous cubic phases in the
mean-field approximation.  For $f_0 < 0$ the excess phases W/O are
stable, for $f_0 > 0$ the microemulsion ME.  Moreover, for
sufficiently negative $g_0$ and not too large $g_2$, a (singly
periodic) lamellar phase $L_{\alpha}$ is found to be stable.
\fig{phi6phasediagram} shows cuts through the parameter space for $g_0
= -3.0$ and $g_0 = -4.5$.  At each point of parameter space, the
elastic moduli can be determined numerically.  In the region where
$\sigma < 0$, the system gains free energy by forming interfaces; the
lamellar phase is stabilized by the short-ranged, repulsive part of
the interaction between them.  In the vicinity of the $\sigma =
0$-line with $\sigma > 0$, the lamellar phase still exists ---
stabilized by the long-ranged, attractive part of the interaction
between interfaces.  However, for a given $g_0$ there is exactly one
point where the phase boundary and the $\sigma = 0$-line touch each
other \cite{a:gozd98}.  Here the interaction between lamellae vanishes
and their distance diverges.  We call this point \emph{unbinding
point} since this divergence resembles a continuous unbinding
transition \cite{a:lipo86,a:leib87}.  We will later make use of the
fact that in the vicinity of the unbinding point the free energy of
the Ginzburg-Landau model (\ref{Phi6FunktionalOrt}) is dominated by
its interfacial contributions.  To our knowledge, no other phases are
stable in the Ginzburg-Landau model.  The (doubly periodic) hexagonal
phase as well as the (triply periodic) cubic phases investigated below
are only metastable.  If one raises $g_0$ to $g_0=-2$, the lamellar
channel vanishes, and microemulsion and excess phases coexist for $f_0
= 0$ and large $g_2$.  As $g_0$ approaches zero from below (with $g_2>0$), 
the lamellar phase disappears.

\subsection{Generating triply periodic structures}
\label{subsec:structures}

\par In order to generate the structures described in
\sec{sec:geometry} as local minima of the Ginzburg-Landau functional, we
construct the Fourier series for the corresponding space group pair
(listed in \tab{nodal}), as described in Sec.~\ref{sec:fourier}, and
terminate it after $N$ stars.  For a given set of parameters
$(f_0,g_0,g_2)$, the free-energy functional (\ref{Phi6FunktionalOrt})
becomes a function of the Fourier amplitudes $A_i$ and $B_i$ $(1 \le i
\le N)$, the constant mode $A_0$ and the lattice constant $a$.
Although the integral in \eq{Phi6FunktionalOrt} could be solved
exactly by using the orthogonality of trigonometric functions, we use
Gaussian integration which is equivalent in efficiency and accuracy,
but easier to implement.  For minimization we use conjugate gradients
and as initial profile the nodal approximations given in \tab{nodal}.
Note that certain structures like P and C(P) need the same Fourier
ansatz but different initial profiles.  In order to obtain initial
profiles for the double structures, we transform the nodal
approximations of the corresponding single structures according to
$\Phi \to 2 \Phi^2 -1$ in real space and then transform these profiles
to Fourier space by using \eq{FourierInvertiert}.  In order to test
for convergence of the Fourier series, we increase $N$ successively
and require both the values for the free-energy density $f$ and the
lattice constant $a$ to level off.  With a typical workstation $N
\approx 100$ is feasible, but $N \approx 30$ turns out to be
sufficient in most cases.  If a profile has converged, the Fourier
amplitudes fall off exponentially for large $|\K_i|$.

\par
\begin{figure}[H]
\begin{center}
\begin{tabular}{cc}
\psfig{file=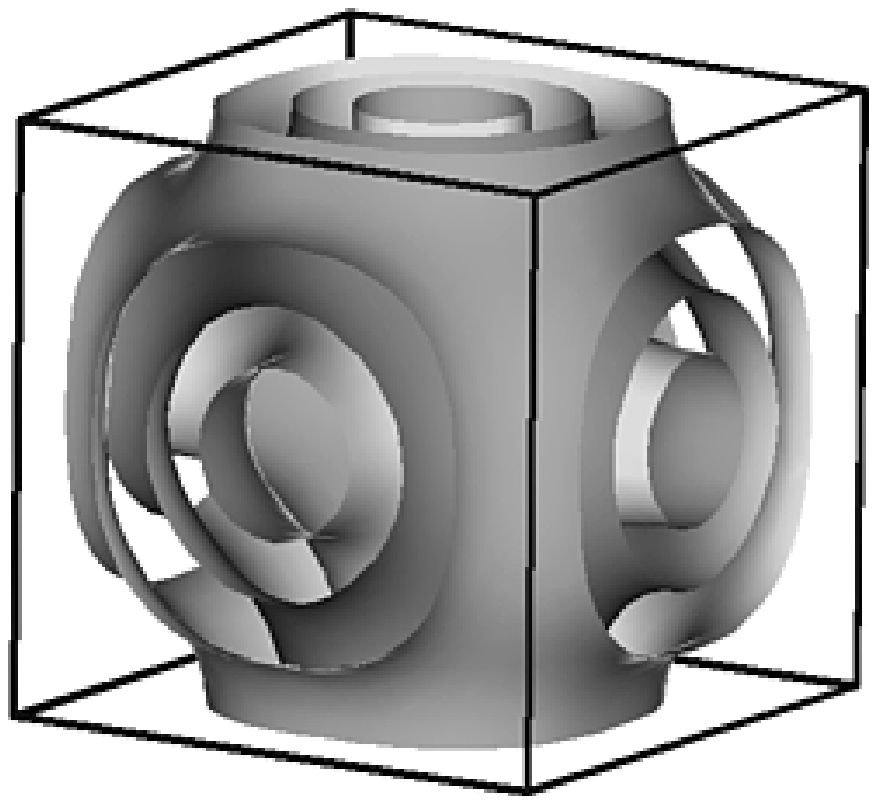,width=7cm} &
\psfig{file=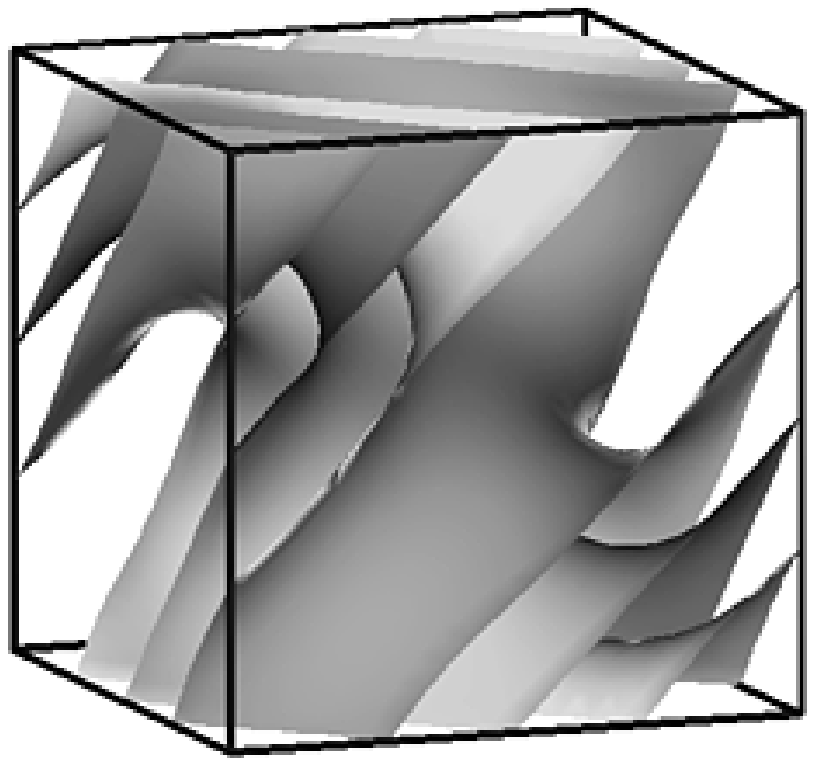,width=7cm} \\
P & D \\
\psfig{file=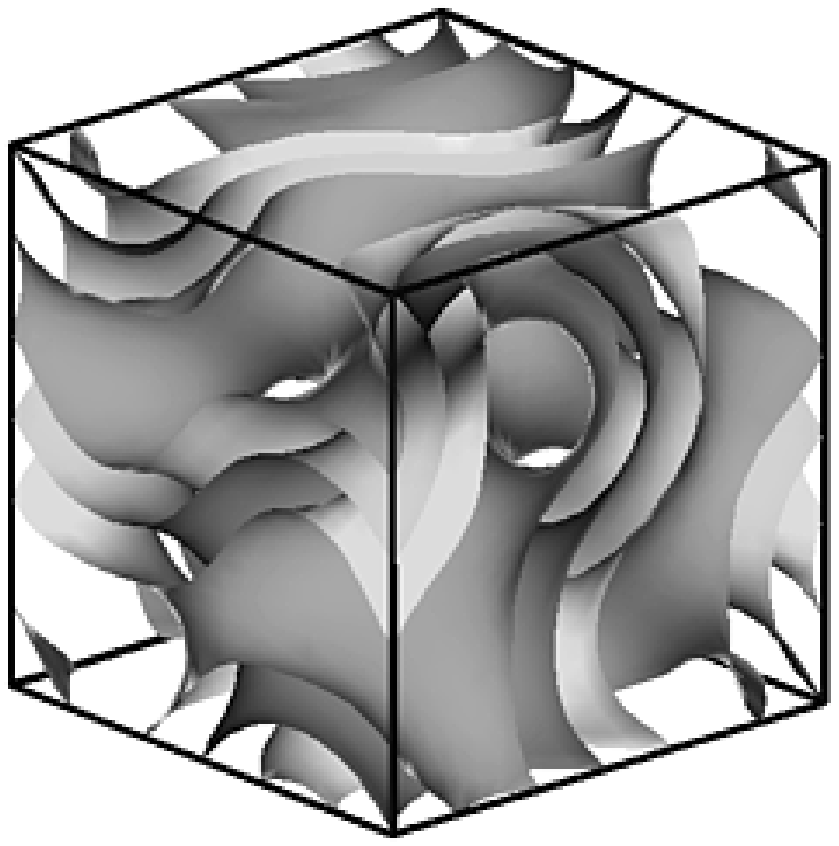,width=7cm} &
\psfig{file=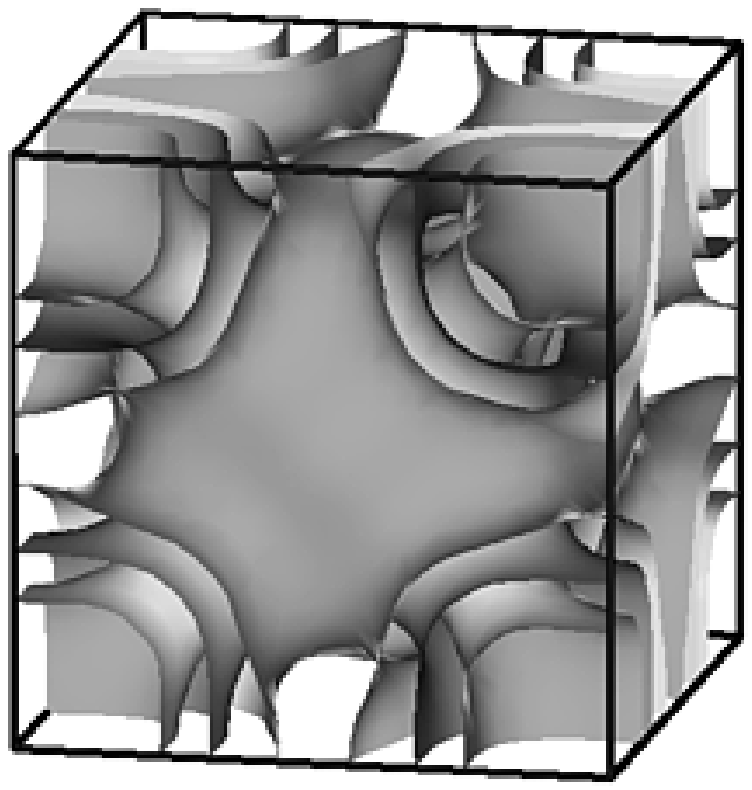,width=7cm} \\
G & I-WP \\
\end{tabular}
\end{center}
\caption{Visualization of single and double structures for P, D, G and
I-WP by their interfaces.  The interfaces of the single structures
appear as middle surfaces to the bilayers of the double structures.
For the single D only one eighth of the unit cell is shown.}
\label{DoubleStructures}
\end{figure}

 We have generated the nine single structures discussed in
\sec{sec:geometry} and the double structures corresponding 
to the four most relevant single structures.  In
\fig{DoubleStructures} we visualize single and double structures of P,
D, G and I-WP by drawing their $\Phi = 0$ iso-surfaces in a
conventional unit cell.  The two interfaces of the double structures
form a bilayer wrapped onto the interfaces of the single structures.
For $f_0 = 0.0$, $g_0 = -3.0$ and $g_2 = 7.01$ our numerical results
are summarized in \tab{Phi6NumericalResults}.  This point is chosen
here in order to compare our results with those of G{\'o}{\'z}d{\'z} and
Ho{\l}yst \cite{a:gozd96b}.  For each structure, we give the free
energy density $f$, the lattice constant $a$ and the volume fraction
of oil $v = 1/V \int d\r (\Phi(\r) + 1)/2$.  The structures are
ordered with respect to their free-energy density $f$; thus, for the
single structures we find the hierarchy G - S - D - I-WP - P etc.  The
scaled surface area $A/a^2$ and the Euler characteristic $\chi$ are
calculated for the conventional unit cell by triangulating the $\Phi =
0$ iso-surface with the marching cube algorithm.  By successively
refining the discretization, we calculate the sum $A_M$ over the
surface areas of the triangles as a function of their number $M$, and
then extrapolate to $A = A_\infty$ by fitting $A_M$ to a linear
$1/M$-dependence.  The values obtained for P, D, G and I-WP turn out
to be very close to the ones for the corresponding minimal surfaces as
calculated from the Weierstrass representations and given in
\tab{WeierstrassTabelle}.  The Euler characteristic $\chi$ can be
calculated as $\chi = C - E/4$ where C is the number of cubes and E is
the number of cube edges cut by the surface.  All our numerical
results agree very well with those reported by G{\'o}{\'z}d{\'z} and
Ho{\l}yst in Ref.~\cite{a:gozd96b} for G, D, P and I-WP.

\par In order to evaluate the curvature properties of the $\Phi =
0$ iso-surfaces, we make use of the fact that --- due to the Fourier
ansatz --- the field $\Phi(\r)$ is known analytically. The mean and
Gaussian curvatures, \cite{a:spiv79,a:barn90}
\begin{equation} \label{isosurface}
H = - \h\ \nabla \cdot \n\ , \quad
K = \h\ \left[(\nabla \cdot \n)^2 
                - \sum_{i,j = 1}^{3}\ \d{i}n_j\ \d{i}n_j \right] \ ,
\end{equation}
where $\n = \nabla \Phi / |\nabla \Phi|$ is the surface normal vector,
can then be calculated exactly.  In \fig{distribution1} we plot the
distribution of $H$ and $K$ over the $\Phi = 0$ iso-surface as
histograms for the structures P, D, G and I-WP in a conventional unit
cell.  Numerical inaccuracies arise here only from the calculation of the
position vectors of the $\Phi = 0$ iso-surface with the marching cube
algorithm, and from the triangle areas $\Delta A_i$, which enter as
weighting factors of $H$ and $K$ for each plaquette of the
triangulation.  The mean curvature is close to zero everywhere.
For the balanced structures P, D and G it is distributed symmetrically
around $H = 0$.  We conclude that the $\Phi = 0$ iso-surfaces of our
numerical solutions are very close, but {\it not} identical to the
corresponding TPMS.  For the Gaussian curvature, we also plot the
distribution obtained from the Weierstrass representation by
numerically evaluating \eq{WeierstrassDarstellungK}.  This is done by
evaluating $K(x + i y)$ on a square lattice in $x$ and $y$, which
covers the fundamental domains for P, D, G and I-WP as given in
\app{sec:weierstrass}, and collect the values in a histogram with
weights $dA(x + i y)$.  The resulting distributions of $K$ are scaled
to unit lattice constant by using the Weierstrass lattice constants
given in \tab{WeierstrassTabelle}, and normalized to $\int dK p(K)=1$.
For P, D and G, these distributions $p(K)$ are identical (apart from
the scale) since they are related to each other by a Bonnet
transformation.  \fig{distribution1} shows that the numerical
distributions agree very well with the exact Weierstrass results.  The
surface area is not very sensitive to the detailed shape of the
interfaces in these cases; the values for the full solutions and the
nodal approximations differ only by a few percent.  Finally, we can
test the procedure used here by employing the Gauss-Bonnet theorem to
calculate the Euler characteristic as $\chi = 1/(2 \pi) \int dA K$.
We find an agreement to three relevant digits.

\par The $K$-distributions of the single structures investigated, for
which {\it no} Weierstrass representations are known, are shown in
\fig{distribution2}.  These surfaces typically have a more complicated
structure within the unit cell, which is reflected in both a larger
number of peaks and in a larger extremal value of the Gaussian
curvature, $K_{min}$. 

\par The Fourier ansatz yields representations of TPMS which are easy
to document and thus straightforward to use in further investigations.
In \tab{improvednodal}, we give improved nodal approximations which
consist of up to six Fourier stars.  When the number of Fourier stars
$N$ is increased, the free-energy density $f$ decreases monotonically,
but the quality of the approximation for the corresponding TPMS does
not necessarily improve in the same way.  Therefore for each
structure we choose an optimal value of $N$.  The quality of the
approximation is judged from the distribution of $H$ and $K$ over the
surface, as described above.  In \tab{comparenodals}, we compare the
nodal approximations from \tab{nodal}, the improved nodal
approximations from \tab{improvednodal} and the exact Weierstrass
representations by monitoring $H_{max}$, the maximum of $|H|$ on the
iso-surface, $\sqrt{\langle H^2 \rangle}$, the square of the variance
of $H$ on the iso-surface, and $K_{min}$, the minimal value of $K$ on
the iso-surface.  For the exact Weierstrass representation, one has
$H_{max} = \sqrt{\langle H^2 \rangle} = 0$.  For the improved nodal
approximations, the values for $H_{max}$ and $\sqrt{\langle H^2
\rangle}$ improve by nearly one order of magnitude when compared with
the nodal approximations.  Also, their distributions of $K$ are very
close to the ones obtained from the Weierstrass representations (the
only exception is G, whose nodal approximation is already quite good,
in particularly for its $K$-distribution).  Thus by adding just a few
more modes, we can considerably improve on the widely used nodal
approximations.  Finally, we want to point out that the values of the
amplitudes for P, D and I-WP given in \tab{improvednodal} are of the
same magnitude as those calculated from the representations for TPMS
obtained in Ref.~\cite{a:ande90}.  The different values arise from the
different shapes of the order parameter profile through the interface.
The Ginzburg-Landau model gives interfaces with a finite width, while
the calculation based on TPMS assumes sharp interfaces.  In fact, the
amplitudes of the Ginzburg-Landau model should approach the
sharp-interface results as the unbinding point is approached.  As the
interface effectively sharpens, the Fourier amplitudes decay more and
more slowly as a function of the wave number $|{\bf K}|$, with a
$|{\bf K}|^{-2}$ behavior in the limit of step-like interfaces --- in
agreement with the well-known $|{\bf K}|^{-4}$ Porod-law for the
scattering intensity of sharp interfaces \cite{a:poro51}.  Note that
\tab{improvednodal} now provides data for G for the first time, which
in fact is the cubic bicontinuous phase most relevant for amphiphilic
systems \cite{a:font90,a:luzz93,a:sedd95}.

\subsection{Hierarchy of structures}
\label{subsec:hierarchy}

\par In order to investigate the relative stability of the cubic
structures in the symmetric Ginzburg-Landau model, we now consider
their interfacial properties.  For a triply-periodic cubic structure,
the free-energy density within the curvature model
(\ref{HelfrichHamiltonian}) reads
\begin{equation} \label{fcurv}
f_{curv} = \frac{1}{a} \left( \sigma A^* \right) +
\frac{1}{a^3} \left( 2 \kappa \int dA \ H^2 
                                     + 2 \bar \kappa \pi \chi \right)
\end{equation}
where $A^*=A/a^2$ is the scaled area per unit cell; the surface area
$A$, the integration of $H^2$ and the Euler characteristic $\chi$
once again refer to the conventional unit cell.  Both terms in
brackets are scale invariant, {\it i.e.}  they do not depend on the
lattice constant $a$.  If the elastic moduli are calculated for the
points in the phase diagram where we minimize the functional, we find
$\sigma < 0$, $\kappa > 0$ and $\bar \kappa < 0$.  In particular, for
$f_0 = 0.0$, $g_0 = -3.0$ and $g_2 = 7.01$ we obtain $\sigma =
-0.84587$, $\kappa = 2.36197$ and $\bar \kappa = -0.97646$.  This
explains why the bicontinuous phases cannot be stable in our model.
The negative surface tension favors modulated phases, and the bending
rigidity favors minimal surfaces; however, only the lamellar phase is
not disfavored by the negative saddle-splay modulus.  We now also
understand why the cubic structures are stable in the Ginzburg-Landau
model with respect to a variation of $a$.  There is a balance between
the negative surface tension term, which favors small values of $a$,
and the positive curvature contributions which favor large $a$.  The
minimization of \eq{fcurv} with respect to $a$ yields
\begin{equation} \label{fmin}
a_{min} = \left( \frac{6 \pi \bar \kappa \chi + 
6 \kappa \int dA H^2}{|\sigma| A^*} \right)^{\h}, \quad
f_{min} = - \left( \frac{4}{27} \right)^{\h}
\left( \frac{(|\sigma| A^*)^3 }{2 \bar \kappa \pi \chi + 
2 \kappa \int dA H^2} \right)^{\h}\ ,
\end{equation}
where $A^*$ is assumed to be (approximately) independent of $a$.
We can now understand why the single structures are found to be so
close to minimal surfaces; the minimum of $\int dA H^2$ (the so-called
\emph{Willmore problem} \cite{a:hsu92}) in this case is $H=0$, which 
in turn minimizes $f_{min}$.  Note that this reasoning is not
rigorous, since the minimization of the free-energy functional with
respect to lattice constant and shape are not independent; the latter
step determines $A/a^2$, which is taken to be constant in the first
step.  

\par We first discuss the single structures without the
``complicated'' phases C(P), C(D), S and C(Y) --- which have larger
lattice constants, a more pronounced modulation within the unit cell,
and stronger interactions between the surfaces. Then we numerically
find $\int dA H^2 \approx 10^{-3}$ (compare \fig{distribution1}), so
that \eq{fmin} becomes
\begin{equation} \label{fgeom}
a_{min} = \left( \frac{6 \pi \bar \kappa \chi}
                         {|\sigma| A^*} \right)^{\h}, \quad
f_{min} = - \left( \frac{4}{27} \right)^{\h}\
\left( \frac{|\sigma|^3}{|\bar \kappa|} \right)^{\h}\ \Gamma
\end{equation}
where $\Gamma = ({A^*}^3 / 2 \pi |\chi|)^{\h} = (A^3 / 2 \pi |\chi|
a^6)^{\h}$ is the \emph{topology index}.  Its exact and numerical
values for the single structures investigated is given in
\tab{WeierstrassTabelle} and \tab{Phi6NumericalResults}, respectively.
This geometrical quantity is independent both of lattice constant and
choice of unit cell.  It can be considered to be a measure for the
porosity of the structure (the larger its value, the less porous) as
well as for the specific surface area (the larger its value, the more
surface area per volume).  Its relevance for amphiphilic systems is
well known \cite{a:ande89,a:hyde89,z:gomp93b}.  In fact it follows
from the isoperimetric relations in three-dimensional space that $A^3
/ |\chi| a^6$ is the only invariant combination of Minkowski
functionals for vanishing integral mean curvature \cite{a:schn93}.

\par A comparison of Eq.~(\ref{fgeom}) with our numerical results from
\tab{Phi6NumericalResults} shows that these formulae systematically
predict a lattice constant and a free-energy density, which are too
large and too low by about $20 \%$, respectively.  However, the
hierarchy of structures as predicted by $f_{min}$ turns out to be
exactly the same as given in \tab{Phi6NumericalResults} by the full
numerical results for $f$:  G - D - I-WP - P - F-RD.  We therefore
conclude that the gyroid structure is the most stable structure since
it has the smallest porosity (the largest topology index).  This
corresponds to the fact that topologically the gyroid's labyrinths
have the smallest connectivity of all structures considered --- G is
the only structure with only three lines meeting at one vertex (D and
P have four and six, respectively).  With the topology index, we have
found a universal geometrical criterion for the relative stability of
the various single bicontinuous cubic phases, which also resolves the
debate about the degeneracy of minimal surfaces in the Canham-Helfrich
approach for vanishing spontaneous curvature.  In order to explain the
non-degeneracy observed experimentally for {\it binary} lipid-water
systems, higher order terms \cite{a:Brui92} and frustration of chain
stretching \cite{a:ande88,a:dues97} have been considered.  In our
description both effects are not necessary.  In fact, in a (balanced)
{\it ternary} system, chain stretching does not provide a plausible
mechanism for lifting the degeneracy of the free energies of different
TPMS, since the presence of oil relieves the frustration in the chain
conformations.  In the part of the phase diagram, where ordered phases
are stable, the free-energy density should contain a {\it negative}
surface tension contribution (compare Refs.~\cite{a:golu89,a:gomp93a})
and the topological term --- which due to the Gauss-Bonnet theorem is
often neglected, but in fact prevents the lattice constant from
shrinking to zero \cite{comm_liko95}.  The same reasoning might
be applied to binary systems by identifying monolayers with bilayers
and oil and water with water I and water II, respectively.  Although
the presence of a negative surface tension is essential in our
argument, we want to emphasize that its magnitude can still be very
small.

\par
\begin{figure}[H]
\begin{center}
\leavevmode 
\psfig{file=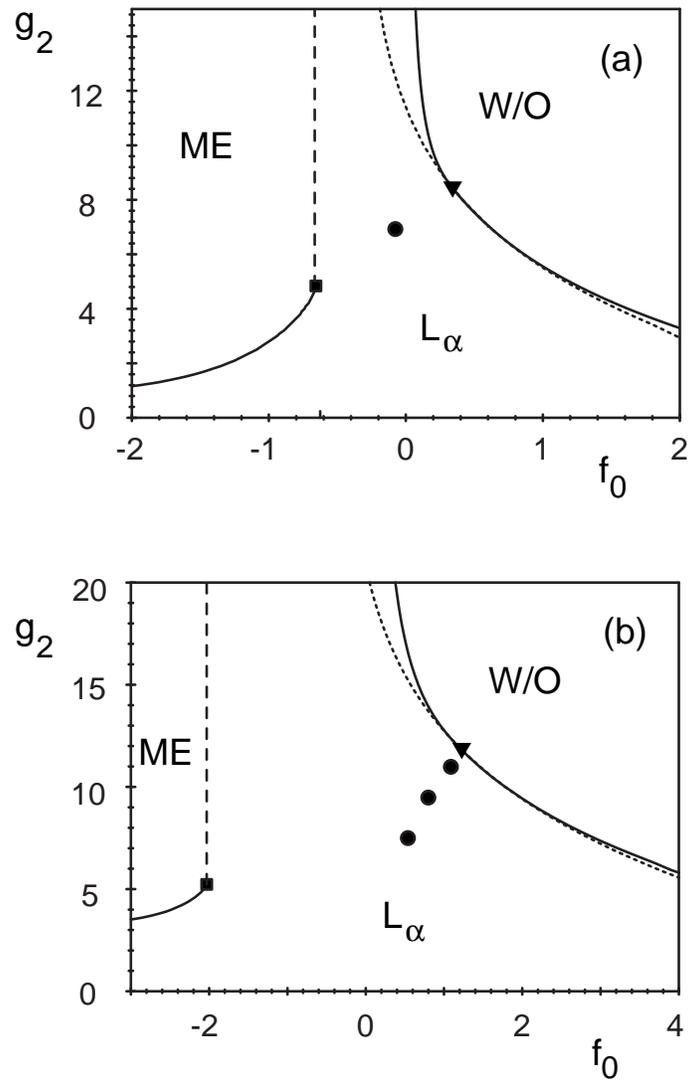,width=9cm}
\end{center}
\caption{Phase diagram for the Ginzburg-Landau model
(\ref{Phi6FunktionalOrt}), (\ref{GL_density}) for (a) $g_0 = -3.0$ and
(b) $g_0 = -4.5$.  Stable phases are the microemulsion ME, the
lamellar phase $L_{\alpha}$ and the water and oil excess phases W and
O.  First and second order transitions are drawn by solid and dashed
lines, respectively.  Tricritical points are marked by squares and
unbinding points by triangles.  The dotted line in the $L_{\alpha}$
stability region is the $\sigma = 0$-line.  Metastable cubic phases are 
calculated at the points marked by full circles.}
\label{phi6phasediagram}
\end{figure}

\begin{figure}[H]
\begin{center}
\leavevmode 
\psfig{file=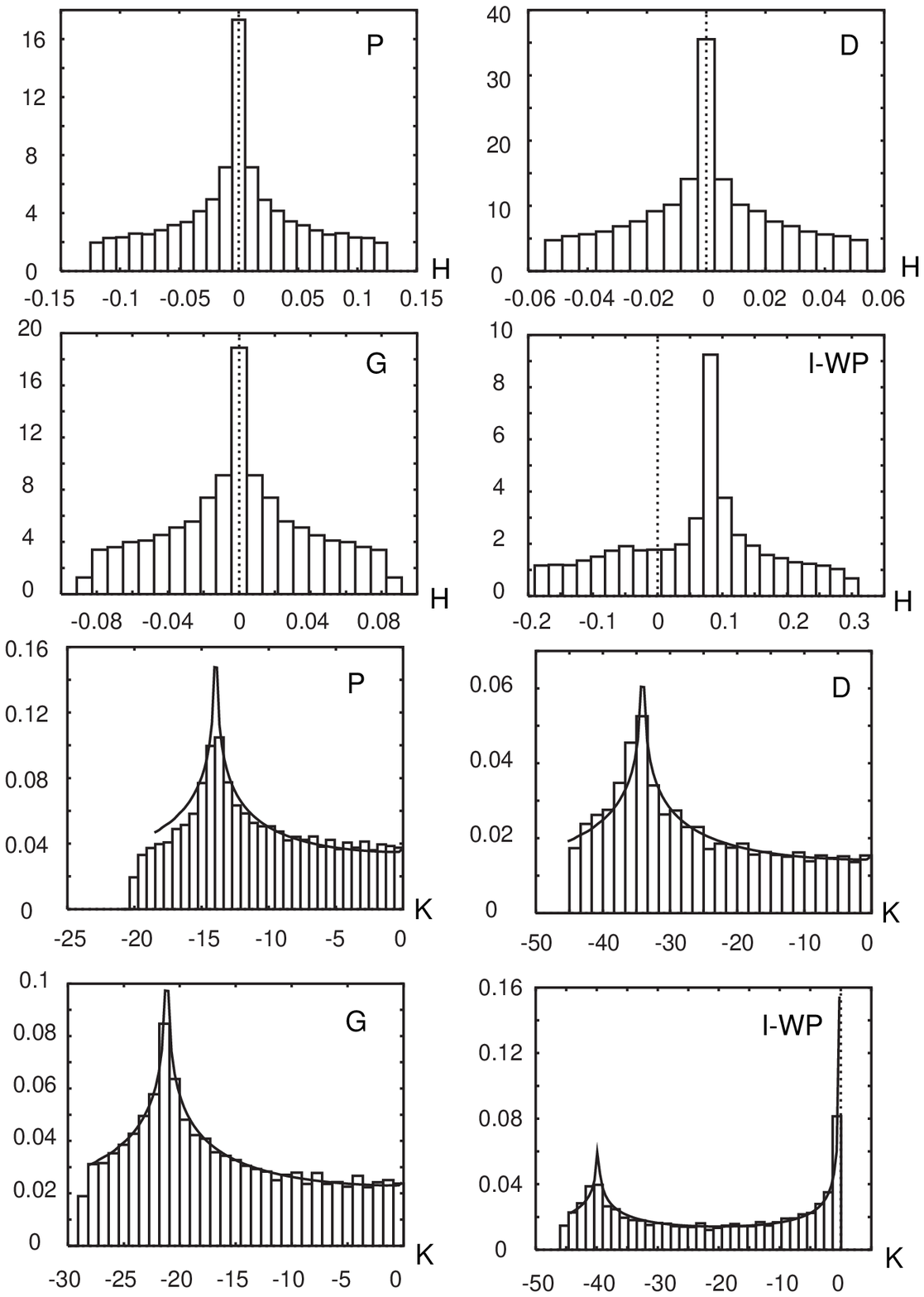,width=15cm}
\end{center}
\caption{Distribution of mean curvature $H$ (top) and Gaussian
curvature $K$ (bottom) in a conventional unit cell, plotted as
histograms for the structures P, D, G and I-WP, for which Weierstrass
representations are known.  Radii of curvature are measured in units
of the lattice constant.  For $K$, we also plot as solid lines the
distributions calculated from the Weierstrass representations.  Due to
the existence of Bonnet transformations between P, D and G, they have
the same shape but different scales in these cases.  The parameters in
the Ginzburg-Landau model are the same as in
Fig.~\ref{phi6phasediagram}a and the $K$-distributions are normalized
to $\int dK p(K)=1$. For these ``simple'' structures, about 30 Fourier 
modes are sufficient.}
\label{distribution1}
\end{figure}

\begin{figure}[H]
\begin{center}
\leavevmode 
\psfig{file=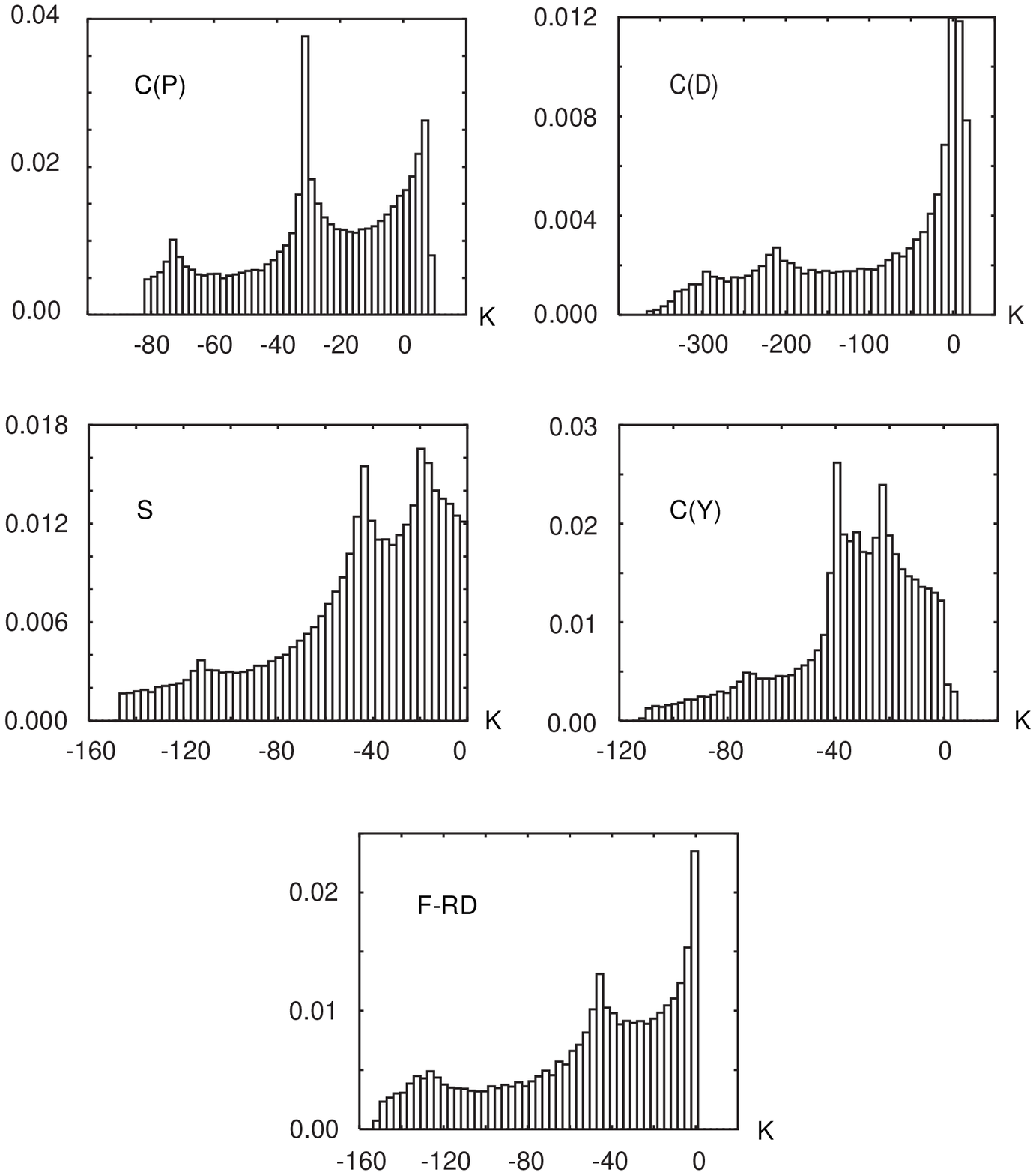}
\end{center}
\caption{Distribution of Gaussian curvature $K$ for single structures,
for which no Weierstrass representations are known.  The normalization
is the same as in Fig.~\ref{distribution1}. These ``complicated''
structures have more pronounced modulation within the unit cell and
larger lattice constants, therefore we use about 50 to 70 Fourier modes
for the numerical minimization.}
\label{distribution2}
\end{figure}

 There are two main reasons why \eq{fgeom} yields the correct
hierarchy in $f_{min}$, but does not reproduce the numerical values
very well.  First, by focussing on the interfacial properties, we have
neglected the contributions to the free-energy density due to direct
interactions between the interfaces, and second, the reasoning leading
to \eq{fmin} is not rigorous.  In order to check whether this is
indeed the origin of the numerical discrepancy, we can make use of the
existence of the unbinding point in the phase diagram,
\fig{phi6phasediagram}.  At this point, both $\sigma$ and the
interaction vanish and the structures investigated should become
exactly minimal surfaces.  However, at the same point the lattice
constant diverges, and both the minimization and the Fourier
ansatz become unfeasible.  Therefore, we study the approach to the
unbinding point along the path shown in \fig{phi6phasediagram}.
Numerically we find for P, D and G that the deviation of the
values for $f$ ($a$) from \eq{fmin} relative to our numerical results
reduces from $14 \%$ ($20 \%$) through $11 \%$ ($16 \%$) to $5 \%$ ($7
\%$) for the three sets of parameters considered.  This demonstrates
the convergence towards minimal surfaces, and verifies our reasoning
above.

\par In contrast to the case of ``simple'' single structures discussed
so far, the topology index $\Gamma$ does {\it not} predict the
hierarchy found numerically for the more complicated single structures
C(P), C(D), S and C(Y), compare \tab{Phi6NumericalResults}.  This may
be due to numerical uncertainties in the value of $\Gamma$, since no
exact results are available for the surface area $A/a^2$.  However,
both the low Ginzburg-Landau free-energy density and the large
topology index of the S-surface indicate that this structure should be
comparable in stability with the G-surface.  We believe that
double structures are not favored in our model because their
interfaces are not minimal surfaces, and can be better described as
surfaces of constant mean curvature.

\par One of the intriguing aspects of the Ginzburg-Landau model is
that it has a very rugged energy landscape with many local minima.  In
fact, we were able to find many more interesting structures, including
more complicated minimal surfaces and surfaces which contain both
saddle-shaped pieces and pieces with positive Gaussian curvature.  The
latter result is related to the observation that we essentially solve
the Willmore problem, which allows for solutions which have both
regions with $K>0$ and regions with $K<0$ --- for example, the
Clifford torus \cite{a:hsu92}.

\section{Scattering intensities and NMR-spectra}
\label{sec:experimental}

\par We expect the nine single and four double structures investigated
in this paper to include all physically relevant bicontinuous cubic
phases in ternary systems near their phase-inversion temperature.  The
representations of these phases can now be used to simulate data for
those experimental techniques, which mainly depend on the geometry of
the structures. This includes small angle scattering (SAS),
transmission electron microscopy (TEM) and nuclear magnetic resonance
(NMR). In particular, our representations can be used to analyze
experimental data for mesoporous systems when bicontinuous cubic
phases have been used as templates \cite{n:atta95}.

\par Since their typical length scales are in the nanometer range,
amphiphilic structures can be investigated by X-ray and neutron small
angle scattering (SAXS and SANS).  The scattering function $S(\k)$ is
obtained from the Fourier transform $\Phi(\k)$ of the density
$\Phi(\r)$ of scattering lengths as $S(\k) = |\Phi(\k)|^2$.  For SANS,
the scattering contrast can be varied by deuterating the sample.  For
balanced single structures, it is clear from above that with bulk
contrast one measures space group $\H$, with film contrast space group
$\G$.  Thus, by varying the contrast in SANS-experiments, the space
group pair of the investigated structure can be determined
\cite{a:raed89,a:radi90}.  With our representations of the cubic
structures, it is straightforward to calculate the corresponding
scattering amplitudes which are needed to analyze the experimental
data.

\par It is quite clear from our results why the program to identify
a structure from its scattering intensity alone is rather
difficult.  For the single structures, the Fourier amplitudes decay
so rapidly with increasing wave vector that the scattering intensity
is dominated by the first peak, while the higher peaks are hardly
detectable.  Even for double structures (or film contrast for single
structures in SANS) the situation hardly improves, since the
scattering intensity also does not feature more than two or three
relevant peaks.  Note that even if the correct space group were
extracted, one still could not be certain about the type of minimal
surface; G and S, for example, have the same space group $Ia\bar3d$
(No.~230) when measured in film contrast.

\par It has been pointed out by Anderson \cite{a:ande90a} that
deuterium NMR is another experimental technique which yields
characteristic fingerprints of bicontinuous cubic phases.  Three
conditions have to be met in this case; first, the amphiphile has to
be deuterated and distributed uniformly over the surface, second, the
sample has to be monocrystalline, and finally, the surfactant sheets
have to be polymerized in order to suppress diffusion within the
surface.  Since deuterium has a nucleus with $I = 1$ and a small
magnetic moment, the quadrupolar interactions dominate and the dipolar
ones can be neglected.  The ${}^2H$-NMR experiments then essentially
measure the distribution $f(x)$ of $x = (3 \cos^2 \rho - 1) / 2$, with
$- \h \le x \le 1$, where $\rho$ is the angle between the external
magnetic field and the amphiphilic director, which corresponds to the
normal vector of the interface \cite{a:seel77}.  In fact, the
electrostatic field at the nucleus does not distinguish between $x$
and $-x$ and the bandshape is $g(x) = f(x) + f(-x)$, with $-1 \le x
\le 1$.

\par For magnetic fields parallel to the (100)-direction, $f(x)$ can 
be calculated exactly for P, D, G and I-WP from their Weierstrass
representations.  The calculation for P, D and G has been performed by
Anderson \cite{a:ande90a}.  We introduce spherical coordinates
$(\theta,\varphi)$ with the polar axis in the direction of the
magnetic field, so that $\rho = \theta$.  The Weierstrass
representation uses the complex $\omega$-plane (see
\app{sec:weierstrass}), which is parametrized in polar coordinates
$(r, \varphi)$.  At every point $\omega$, the normal vector follows by
stereographic projection to the unit sphere. \eq{StereographicProjection} 
then gives the one-to-one correspondence
\begin{equation} \label{NMRx}
x = \frac{r^4 - 4 r^2 + 1}{(1 + r^2)^2}\ .
\end{equation}
With $r' = r(x')$ and \eq{WeierstrassDarstellungK}, the distribution
becomes
\begin{eqnarray}  
f(x') &=& \int_{0}^{\infty} r\ dr\ \int_{0}^{2 \pi} d\varphi\ 
|R(r,\varphi)|^2\ (1 + r^2)^2\ \delta(x(r) - x')   \nonumber \\
  &=& \int_{0}^{\infty} r\ dr\ \int_{0}^{2 \pi} d\varphi\ 
       |R(r,\varphi)|^2\ (1 + r^2)^2\ 
 \left|\frac{\partial x}{\partial r}({r'})\right|^{-1} \delta(r - {r'}) 
                                                   \nonumber \\
&=& \frac{(1 + {r'}^2)^5}{12\ |{r'}^2 - 1|}\ 
              \int_{0}^{2 \pi} d\varphi\ |R({r'},\varphi)|^2
\label{NMRfx}           
\end{eqnarray}
where $R$ is the generating function given for P, D, G and I-WP in
\app{sec:weierstrass}. We denote the remaining integral by $I(r')$. 
P, D and G yield the same result \cite{a:ande90a},
\begin{equation} \label{NMRPDG}
I_{PDG}(r) = \frac{32\ K(k)}{r^4 \sqrt{(\epsilon_1 - 1) 
                                          (\epsilon_2 + 1)}} \ ,
\end{equation}
since they are related by a Bonnet transformation. Here, $K(k)$ is the 
complete elliptical integral of the first kind, and
\begin{equation} \label{NMRPDGParameter}
\epsilon_1 = \frac{a_+ + a_- r^8}{r^4} \ , \ \ \ 
\epsilon_2 = \frac{a_- + a_+ r^8}{r^4}  \ , \ \ \ 
a_{\pm} = \frac{7 \pm \sqrt{48}}{2}  \ , \ \ \ 
k^2 = \frac{2 (\epsilon_1 - \epsilon_2)}{(\epsilon_1 - 1) 
                                            (\epsilon_2 + 1)}\ .
\end{equation} 
For I-WP, we find from Eq.~(\ref{NMRfx})
\begin{equation} \label{NMRIWP}
I_{I-WP}(r) = \frac{32 \pi}{(r^2 - r^{10})^{\frac{2}{3}}}\ 
P_{- \frac{1}{3}}\left(\frac{1 + r^8}{1 - r^8}\right)
\end{equation} 
where $P_{\nu}(x)$ is a Legendre function of the first kind.  Two values
of $r$ correspond to any given $x$ in \eq{NMRx}, $0 \le r \le 1$ for
the southern hemisphere and $1 \le r \le \infty$ for the northern
hemisphere.  However, for the direction and the structures considered
both $x$ and $dA$ do not change when the northern is mapped onto the
southern hemisphere by a reflection through the $xy$-plane (this can
be verified by inverting the complex plane in the unit circle).  It is
therefore sufficient to evaluate \eq{NMRfx} for $0 \le r \le 1$.  Then
the bandshapes $g(x) = f(x) + f(-x)$ for both P, D, G and I-WP are
found to have peaks at $x = \pm \h$, which are already present in the
{\it Pake pattern}, $f(x) = 2 \pi / \sqrt{6 x + 3}$, of isotropic
samples.  However, additional peaks appear at $x = 0$ for P, D and G
and at $x = \pm 1$ for I-WP, which correspond to the positions of the
flat points ((111) for P, D and G, and (100) for I-WP).

\par
\begin{figure}[H]
\begin{center}
\leavevmode 
\psfig{file=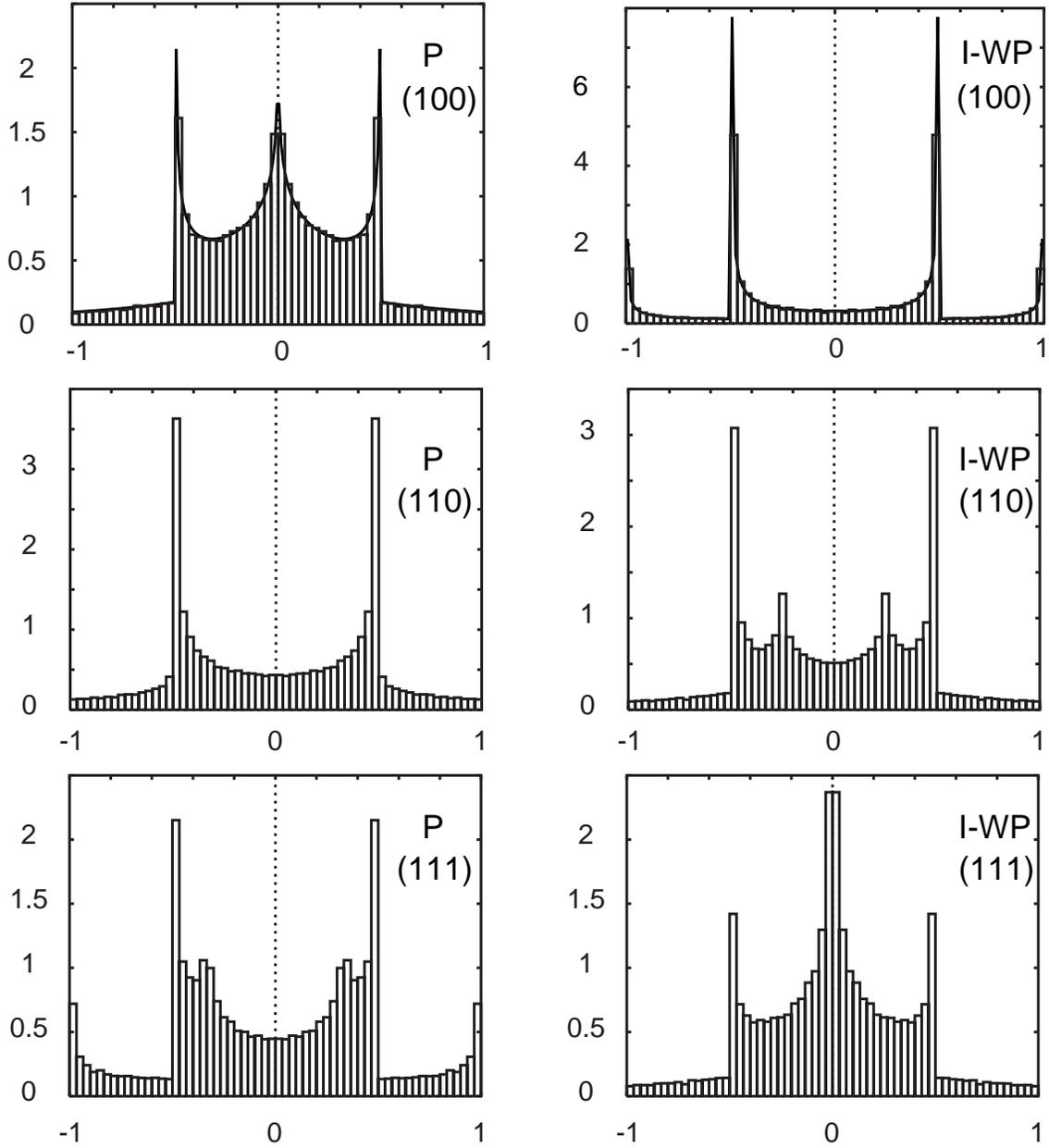,width=15cm}
\end{center}
\caption{${}^2H$-bandshapes for the structures P and I-WP and for the
three high-symmetry directions $(100)$, $(110)$ and $(111)$.  Results
from the Weierstrass representations for the $(100)$-direction are 
drawn as solid lines.}
\label{nmr1}
\end{figure}

\begin{figure}[H]
\begin{center}
\leavevmode 
\psfig{file=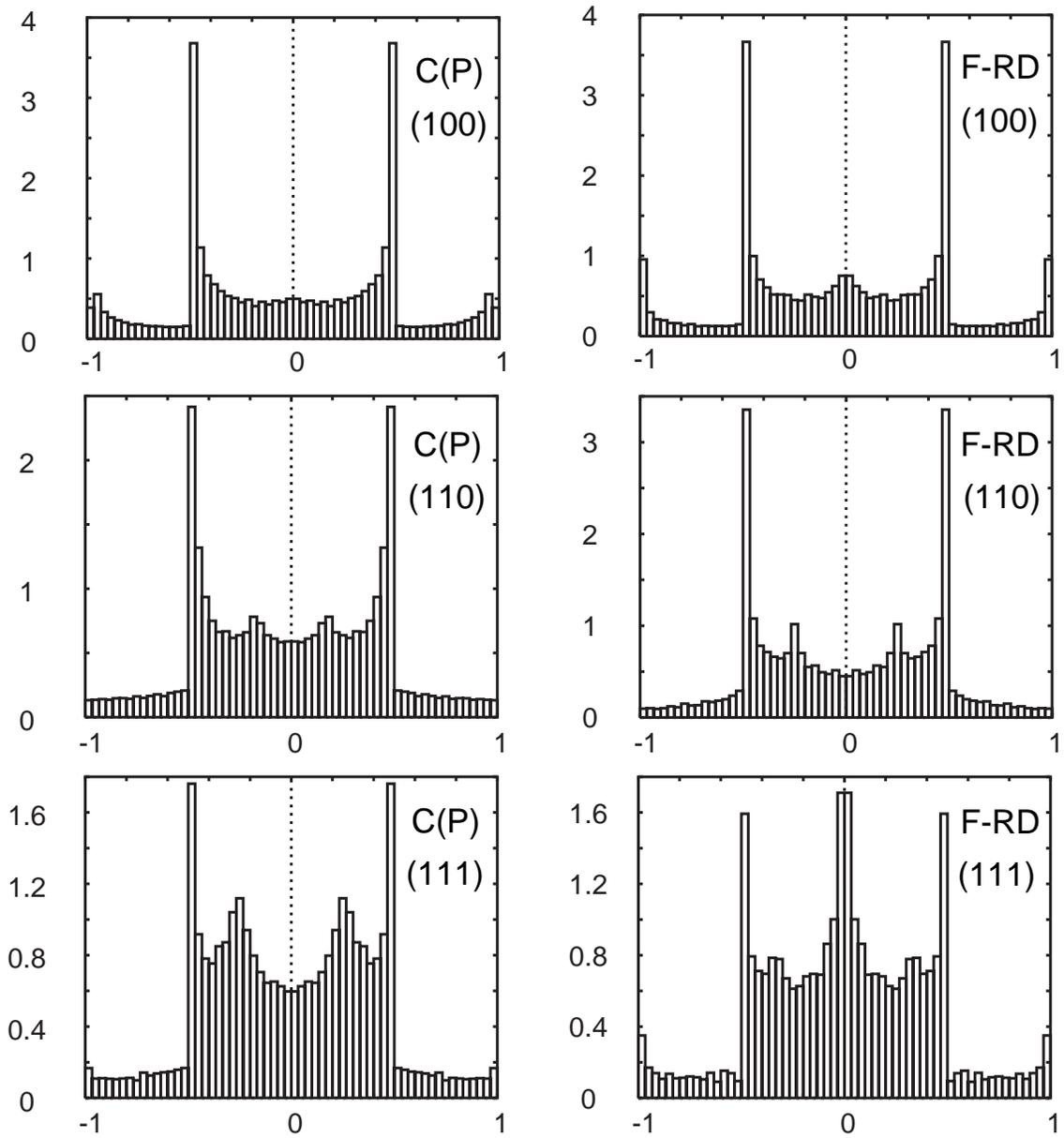,width=15cm}
\end{center}
\caption{${}^2H$-bandshapes for the structures C(P) and F-RD and for the
three high-symmetry directions $(100)$, $(110)$ and $(111)$.}
\label{nmr2}
\end{figure}

 For the same structures, but other directions, and for
other structures, an exact calculation is not possible.  However, we 
can numerically determine the distribution $f(x)$ for any direction and 
any structure in the same way used above to calculate the distribution of 
$H$ and $K$ over the surface.  In fact, with the Fourier ansatz, this can
be done with much better resolution than for a real-space representation
\cite{a:gozd97}.  In \fig{nmr1} we show our results  for P and I-WP and
the three high-symmetry directions.  For the (100)-direction, we also
show the exact results following from \eq{NMRfx} with \eq{NMRPDG} and
\eq{NMRIWP} for P and I-WP, respectively; the agreement with the
numerical results is excellent. In \fig{nmr2} our results are shown
for C(P) and F-RD for which no Weierstrass representations are
known.

\section{Summary}
\label{sec:conclusion}

\par We have presented here a systematic investigation of bicontinuous
cubic phases in ternary amphiphilic systems.  In these structures the
amphiphilic monolayers form triply periodic surfaces with cubic
symmetry.  We distinguish between single and double structures which
are characterized by one or two monolayers, respectively.  In order to
further classify the structures within each of these groups, we used
the crystallographic classification of triply periodic minimal
surfaces by Fischer and Koch.  Thus for single structures we
distinguish between balanced and non-balanced structures; the first
class is further subdivided into spanning and non-spanning structures.
Finally we end up with a list of nine single structures of interest
(P, C(P), D, C(D), S, C(Y), G, I-WP, F-RD).  For each of these there
exists a corresponding double structure from which we considered those
corresponding to the simple and physically relevant single structures
P, D, G and I-WP.

\par In the framework of a Ginzburg-Landau model for ternary
amphiphilic systems, we generated the nine single and four double
structures using the Fourier ansatz and the theories of space groups
and color symmetries.  Compared to real-space minimization, the
Fourier approach has the advantage of efficient numerics and easy
documentation.  For P, D, G and I-WP, we gave improved nodal
approximations which give much better approximations to triply
periodic minimal surfaces than the widely used nodal approximations by
von Schnering and Nesper.  We showed that the free-energy density of the
single structures can be calculated from an effective surface
Hamiltonian with negative surface tension, positive bending rigidity
and negative saddle-splay modulus.  Due to the water-oil symmetry of
the model, the spontaneous curvature vanishes and structures with zero
mean curvature are favored.  A comparison with the exact Weierstrass
representations for P, D, G and I-WP shows that the single structures
can be made to closely approach triply periodic minimal surfaces by
appropriately tuning the model parameters.  For vanishing mean
curvature term, their relative stability is determined by the topology
index $\Gamma = (A^3 / 2 \pi |\chi| a^6)^{\h}$.  This explains the
hierarchy G - D - I-WP - P.  Thus for water-oil symmetry the single
gyroid is the most stable cubic bicontinuous phase since it has the
smallest porosity.

\par The representations obtained for both single and double
structures can now be used for further physical investigations.  We
have employed them to calculate the distribution of the Gaussian
curvature for C(P), C(D), S, C(Y) and F-RD, which were not known
before.  Furthermore, we have determined several quantities which can
be measured in SAS- and ${}^2H$-NMR-experiments.

\par In our Ginzburg-Landau model for balanced ternary systems,
modulated phases are favored whose interfaces are minimal surfaces.
However, only the lamellar phase is stable, since the bicontinuous
phases are disfavored by the {\it negative saddle-splay modulus}.
Thus, other energetic contributions have to be considered to stabilize
bicontinuous phases.  This can be long-ranged interactions (like the
van-der-Waals or electrostatic interactions) or higher order terms in
the curvature energy.  Assuming that these contributions typically
have a similar effect on all bicontinuous phases, we then expect the
gyroid phase G to be most prominent.  One of the intriguing conclusion
of our calculations is the large relative stability of the {\it
S-surface}, as indicated by its low free-energy density in the
Ginzburg-Landau model and by its large topology index.  Thus, for
single cubic phases in balanced ternary amphiphilic systems, the
S-surface could be a possible alternative to the G-surface, whose
double version seems to be so ubiquitous in lipid-water mixtures.

\par We want to emphasize that bicontinuous phases can of course be
induced by a {\it positive saddle-splay modulus}.  However, in this
case the lattice constant should be on the order of the size of the
amphiphilic molecules, and the application of the curvature energy
model becomes questionable.  This may indeed be the case in many
lipid-water systems, where the amphiphile volume fraction in the
bicontinuous cubic phases is larger than $50\%$.

\par\noindent
{\large \bf Acknowledgments} \\ 
We thank W.~G{\'o}{\'z}d{\'z} and C.~Burger for many helpful discussions.

\newpage

\appendix
\section{Weierstrass representations}
\label{sec:weierstrass}

\par Exact representations are known for four cubic TPMS
\cite{a:nits89,a:dier92,a:fodg92b,a:fodg92a}.  Consider the composite
mapping of the surface into the complex plane, which consists of two
parts; first, each point of the surface is mapped to a point on unit
sphere which is defined by the normal vector on the surface, then the
unit sphere is mapped into the complex plane by stereographic
projection.  The \emph{Weierstrass representation formulae} invert
this mapping and therefore map certain complicated regions $\Omega
\subset {\Bbb C}$ of the complex plane onto a fundamental piece of the
TPMS,
\begin{equation} \label{WeierstrassDarstellung}
\f(x,y) = Re \int_0^{\omega} dz\ R(z)\ 
{\left( \begin{array}{c} 1 - z^2 \\ i (1 + z^2) \\ 2 z 
                                           \end{array} \right)}\ .
\end{equation}
where $\omega = x + i\ y$.  The replication of this surface segment
with the symmetries of the space group $\H$ of the oriented surface
then gives the whole TPMS.  It can be shown that each meromorphic
function $R(\omega)$ corresponds to a minimal surface.  However, only
few meromorphic functions lead to embedded TPMS.  For the D-surface,
one has $R(\omega) = (\omega^8 - 14 \omega^4 + 1)^{-\frac{1}{2}}$, and
$\Omega$ is the common area of the four circles with radii $\sqrt{2}$
around the points $(\pm 1 \pm i)/\sqrt{2}$.  The Bonnet transformation
$R(\omega) \to e^{i \theta} R(\omega)$ with $\theta = \pi/2$
transforms the D-surface piece into one for the P-surface, and $\theta
= 38.015^o$ does the same for the G-surface.  The generating function
for I-WP was found only recently to be $R(\omega) = (\omega (\omega^4
+ 1))^{-\frac{2}{3}}$ \cite{a:lidi90,a:cvij94}.  Here $\Omega$ is the
common area of the circle with radius $\sqrt{2}$ around the point $(1
+ i)/\sqrt{2}$ with the unit circle, the lower half plane and the
lower half plane rotated anti-clockwise by $\pi / 4$. 

\par Given a generating function $R(\omega)$ for a certain TPMS,
several of its properties can be readily calculated.  At a given point
$\omega = x + i y = r \exp(i \varphi)$, the normal are determined
by stereographic projection from the complex plane to the unit sphere,
\begin{equation} \label{StereographicProjection}
\cos \theta = \frac{r^2 - 1}{r^2 + 1} 
\end{equation}
in spherical coordinates $(\theta, \varphi)$. Gaussian curvature and 
differential area element follow as
\begin{equation} \label{WeierstrassDarstellungK}
K(\omega) = \frac{-4}{|R(\omega)|^2 (1 + |\omega|^2)^4}\ , \quad 
dA(\omega) = |R(\omega)|^2\ (1 + |\omega|^2)^2\ dx dy\ . 
\end{equation}
Therefore the poles of $R(\omega)$ correspond to the (isolated) flat
points of the surface (that is points with $K = 0$).  Since $K$ and
$dA$ depend only on the modulus of $R$, surfaces related by a Bonnet
transformation have the same distribution of $K$ and the same surface
area for one fundamental domain.  \tab{WeierstrassTabelle} shows data
which are obtained from the Weierstrass representations.

\newpage

\begin{table}[H]
\begin{center}
\begin{tabular}{|c|c|c|c|c|c|c|} \hline
     & $N_d$ & $\chi$ & $a$ & $A^*$  & $\Gamma$ & $v$ \\ \hline
P    & 12  & -4 & 2.1565 & 2.345103 & 0.716346 & 0.5 \\ \hline
D    & 48  & -16 & 3.3715 & 3.837785 & 0.749844 & 0.5 \\ \hline
G    & 24  & -8 & 2.6562 & 3.091444 & 0.766668 & 0.5 \\ \hline
I-WP & 48  & -12 & 7.9499 & 3.464102 & 0.742515 & 0.536 \\ \hline
\end{tabular}
\end{center}
\caption{Properties of the four cubic triply periodic minimal surfaces
G, D, I-WP and P, for which (exact) Weierstrass representations are
known.  $N_d$ is the number of copies of the fundamental domain
which are necessary to build up the surface in one conventional unit
cell.  $\chi$ is the Euler characteristic, $a$ the lattice constant,
and $A^*=A / a^2$ the scaled surface area in the conventional unit cell.
$\Gamma = (A^3 / 2 \pi | \chi| a^6)^{\h}$ is the topology index and
$v$ the volume fraction of one of the two labyrinths.}
\label{WeierstrassTabelle}
\end{table}

\begin{table}[H]
\begin{center}
\begin{tabular}{|l|l|l|l|l|} \hline
& $\H$ & $\G$ & $\H \to \G$ & nodal approximations \\ \hline \hline
P & $Pm\bar{3}m$ (221) & $Im\bar{3}m$ (229) & $P \to I$ & 
Eccc(100) \\ \hline 
C(P) & $Pm\bar{3}m$ (221) & $Im\bar{3}m$ (229) & $P \to I$ & 
Eccc(100) + Eccc(111) \\ \hline 
D & $Fd\bar{3}m$ (227) & $Pn\bar{3}m$ (224) & $F \to P$ & 
(Eccc + Ecss + Escs + Essc)(111) \\ \hline 
C(D) & $Fd\bar{3}m$ (227) & $Pn\bar{3}m$ (224) & $F \to P$ & 
(Eccc + Ecss - Escs - Essc)(311) \\ \hline 
S & $I\bar{4}3d$ (220) & $Ia\bar{3}d$ (230) & $\bar43m \to m\bar3m$ & 
(Ecsc + Occs)(211) \\ \hline
C(Y) & $P4_332$ (212) & $I4_132$ (214) & $P \to I$ & 
- (Eccc + Osss + Esss + Occc)(111) \\
& & & & + 6\ (Essc + Oscc + Eccs + Ocss)(210) \\ \hline \hline 
G & $I4_132$ (214) & $Ia\bar{3}d$ (230) & $432 \to m\bar3m$ & 
(Escc + Ocsc)(110) \\ \hline \hline 
I-WP & $Im\bar{3}m$ (229) & & & 
2\ Eccc(110) - Eccc(200) \\ \hline 
F-RD & $Fm\bar{3}m$ (225) & & & 
4\ Eccc(111) - 3\ Eccc(220) \\ \hline
\end{tabular}
\end{center}
\caption{Group - subgroup pairs $\G - \H$, the relation $\H \to \G$
between them and nodal approximations for the cubic minimal surfaces.
They are grouped into three classes: spanning with three-dimensional
nets, balanced but non-spanning, and non-balanced.  $\G$ and $\H$ for
balanced single structures differ either in Bravais lattice or in
point group (but not in both).  Non-balanced single structures have
$\H \equiv \G$.  The Fourier ansatz has to be chosen as follows: $\H$
with color symmetry (given by $\H \to \G$) for single structure, and
$\G$ for double structure.  Nodal approximations only consider the
space group information given by $\G - \H$.  We use the definitions
$Epqr = p(h x) q(k y) r(l z) + p(h y) q(k z) r(l x) + p(h z) q(k x)
r(l y)$ and $Opqr = p(h x) q(k z) r(l y) + p(h y) q(k x) r(l z) + p(h
z) q(k y) r(l x)$, where $p$, $q$ and $r$ can be either $c$ (cosine)
or $s$ (sine).}
\label{nodal}
\end{table}

\begin{table}[H]
\begin{center}
\begin{tabular}{|l|l|l|l|l|l|l|} \hline
     & $f$ & $a$ & $v$ & $A^*$ & $\chi$ & $\Gamma$ \\ \hline
G    & -0.19096 & 10.08 & 0.500 & 3.09140 & -8 & 0.76665 \\ \hline
S    & -0.18962 & 17.72 & 0.500 & 5.41454 & -40 & 0.79474 \\ \hline
D    & -0.18870 & 12.56 & 0.500 & 3.83755 & -16 & 0.74978 \\ \hline
I-WP & -0.18112 & 11.82 & 0.527 & 3.46367 & -12 & 0.74238 \\ \hline
P    & -0.18109 &  7.89 & 0.500 & 2.34516 & -4 & 0.71637 \\ \hline
C(Y) & -0.18061 & 14.92 & 0.500 & 4.46108 & -24 & 0.76730 \\ \hline
C(D) & -0.17382 & 28.93 & 0.500 & 8.25578 & -144 & 0.78862 \\ \hline
F-RD & -0.16311 & 17.36 & 0.532 & 4.75564 & -40 & 0.65417 \\ \hline
C(P) & -0.16239 & 14.13 & 0.500 & 3.80938 & -16 & 0.74154 \\ \hline \hline
GG   & -0.18850 & 18.47 & 0.495 & 5.33712 & -16 & \\ \hline
DD   & -0.18549 & 11.59 & 0.502 & 3.30239 & -4 & \\ \hline
PP   & -0.17700 & 14.98 & 0.523 & 4.09739 & -8 & \\ \hline
IWP2 & -0.16193 & 21.55 & 0.532 & 6.29274 & -24 & \\ \hline
\end{tabular}
\end{center}
\caption{Properties of cubic bicontinuous phases as calculated
numerically from the Ginzburg-Landau functional at the point $f_0 =
0.0$, $g_0 = -3.0$ and $g_2 = 7.01$.  The lamellar phase is stable at
this point with $f = -0.20807$, the hexagonal phase metastable with $f
= -0.17552$.  The nine single (top) and four double (bottom) cubic
structures are ordered with respect to their free-energy density $f$.
$a$ is the lattice constant, $v$ the volume fraction of oil.  The
values for scaled surface area $A^*=A/a^2$ and Euler characteristic
$\chi$ are given for the conventional unit cell.  From
these numbers the topology index $\Gamma$ can be calculated.  Note
that the symmetry between water and oil is broken not only for
non-balanced, but also for double structures. At this point of the
phase diagram, about 50 Fourier modes are sufficient and have been 
used for all structures.}
\label{Phi6NumericalResults}
\end{table}

\begin{table}[H]
\begin{center}
\begin{tabular}{|l|l|l|l|} \hline
& mode & function & amplitude \\ \hline \hline
P       & (1 0 0) & Eccc + Occc &  0.2260 \\
        & (1 1 1) & Eccc + Occc & -0.0516 \\
        & (2 1 0) & Eccc + Occc & -0.0196 \\
        & (3 0 0) & Eccc + Occc & -0.0027 \\ \hline
D       & (1 1 1) & Eccc + Ecss + Escs + Essc 
                  + Occc + Ocss + Oscs + Ossc &  0.1407 \\
        & (3 1 1) & Eccc + Ecss - Escs - Essc 
                  + Occc + Ocss - Oscs - Ossc &  0.0209 \\
        & (3 1 3) & Eccc - Ecss + Escs - Essc 
                  + Occc - Ocss + Oscs - Ossc & -0.0138 \\
        & (1 1 5) & Eccc + Ecss + Escs + Essc 
                  + Occc + Ocss + Oscs + Ossc & -0.0028 \\
        & (3 3 3) & Eccc + Ecss + Escs + Essc 
                  + Occc + Ocss + Oscs + Ossc & -0.0021 \\
        & (3 1 5) & Eccc + Ecss - Escs - Essc 
                  + Occc + Ocss - Oscs - Ossc &  0.0011 \\ \hline
G       & (1 1 0) & Escc + Ocsc &  0.3435 \\
        & (0 3 1) & Ecsc - Occs & -0.0202 \\
        & (2 2 2) & Esss + Osss & -0.0106 \\
        & (2 1 3) & Ecsc + Occs &  0.0298 \\
        & (3 0 3) & Eccs + Oscc &  0.0016 \\
        & (1 1 4) & Escc + Ocsc & -0.0011 \\ \hline
I-WP    & (0 0 0) & 1    &  0.0652 \\
        & (1 1 0) & Eccc &  0.5739 \\
        & (2 0 0) & Eccc & -0.1712 \\
        & (2 1 1) & Eccc & -0.1314 \\
        & (2 2 0) & Eccc &  0.0184 \\ \hline
\end{tabular}
\end{center}
\caption{Improved nodal approximations for the four most relevant
cubic structures P, D, G and I-WP.  In contrast to the nodal
approximations given in \tab{nodal}, the extremal values of the
functions $\Phi({\bf r})$ are close to $\pm 1$, since they are
calculated from the Ginzburg-Landau model (\ref{Phi6FunktionalOrt}), 
(\ref{GL_density}). $Epqr$ and $Opqr$, with $p, q, r \in \{c,s\}$, 
are defined in \tab{nodal}.}
\label{improvednodal}
\end{table}

\begin{table}[H]
\begin{center}
\begin{tabular}{|l|l|l|c|c|l|c|c|} \hline
& \multicolumn{3}{c|}{Nodal} & \multicolumn{3}{c|}{Improved nodal} 
                                             & Weierstrass \\ \hline
& $H_{max}$ & $\sqrt{\langle H^2 \rangle}$ & $K_{min}$ 
& $H_{max}$ & $\sqrt{\langle H^2 \rangle}$ & $K_{min}$ & $K_{min}$ \\ \hline
P & 1.282 & 0.579 & -39.48 & 0.157 & 0.071 & -16.84 & -18.51 \\ \hline
D & 0.384 & 0.184 & -39.48 & 0.048 & 0.022 & -46.22 & -45.24 \\ \hline
G & 0.208 & 0.102 & -29.60 & 0.078 & 0.046 & -29.42 & -28.08 \\ \hline
I-WP & 3.224 & 1.326 & -87.34 & 0.273 & 0.113 & -45.47 & -44.37 \\ \hline
\end{tabular}
\end{center}
\caption{Comparison of nodal approximations, improved nodal
approximations and exact Weierstrass representations.  $H_{max}$ is
the maximum of $|H|$, $K_{min}$ the minimum of $K$, and $\langle H^2
\rangle = 1/A \int dA H^2$ the variance of $H$ over the surface.  The
form of nodal and improved nodal approximations is listed in
\tab{nodal} and \tab{improvednodal}, respectively.}
\label{comparenodals}
\end{table}

\newpage

\end{document}